%% file: paper.tex
\documentclass[11pt]{article}

\usepackage[final]{acl}

\usepackage[utf8]{inputenc}
\usepackage[T1]{fontenc}
\usepackage{times}
\usepackage{latexsym}
\usepackage{graphicx}
\usepackage{booktabs}
\usepackage{array}
\usepackage{amsmath,amssymb}
\usepackage{xcolor}
\usepackage{url}
\title{\textbf{A Factorial Ablation of a Speech-to-SFT Pipeline:\\
Differential Effects on Data Quality and Downstream Transfer}}
\author{Wonsup Shin \\
  Flitto \\
  \texttt{wonsup.shin@flitto.com} \\\And
  Jingu Kim \\
  Flitto \\
  \texttt{jin@flitto.com}}
\date{}

\begin{document}
\maketitle

\begin{abstract}
\noindent
Industry pipelines that turn speech into supervised fine-tuning (SFT) data via multi-stage refinement are increasingly adopted but, to our knowledge, have not been publicly ablated stage-by-stage, leaving each stage's marginal value unknown. We design a production-ready speech-to-SFT pipeline in which transcript refinement (Phase~0) and SFT data quality refinement (Phase~2) are independently toggleable, yielding a $2\times 2$ factorial design. For each condition, we generate QA-form SFT data from Korean medical and finance conference recordings and fine-tune 9 models (5 LLM families, 2.4B--70B); we evaluate with four cross-provider LLM judges, a blind six-expert human evaluation, and three downstream MCQA benchmarks. Our central finding is that, under a fixed, standard SFT recipe, improvements in QA data quality do not transfer uniformly into downstream MCQA gains. 4-judge quality rises consistently, yet the cross-model mean MCQA gain is not significant; positive transfer concentrates on family$\leftrightarrow$domain-aligned pairs. All six human raters report higher full-pipeline quality, confirming the LLM-judge direction. An STT-engine swap to Whisper-medium confirms pipeline robustness. An LLM-difficulty audit shows the two frontier LLMs admit unknown on ${\sim}8\%$ of QA on average; we release samples, prompts, code, and all SFT checkpoints.
\end{abstract}

\section{Introduction}
\label{sec:intro}

Industry pipelines that turn in-house speech assets into supervised fine-tuning (SFT) data are increasingly adopted but rarely reported in the open literature. Such pipelines combine speech-to-text (STT), transcript refinement, SFT data generation, and SFT data quality refinement, yet, to our knowledge, no public study performs a per-stage ablation of these refinement stages, leaving each stage's marginal value unknown to practitioners. Even prior speech-to-SFT studies have targeted general capabilities~\citep{pan2024cosmic, amazon2025sift} or applied non-SFT methods such as continuous pre-training~\citep{li2025podgpt}, so per-stage ablation on a domain-specific case remains underexplored.

\begin{figure*}[t]
  \centering
  \includegraphics[width=0.95\linewidth, trim=0 30 0 80, clip]{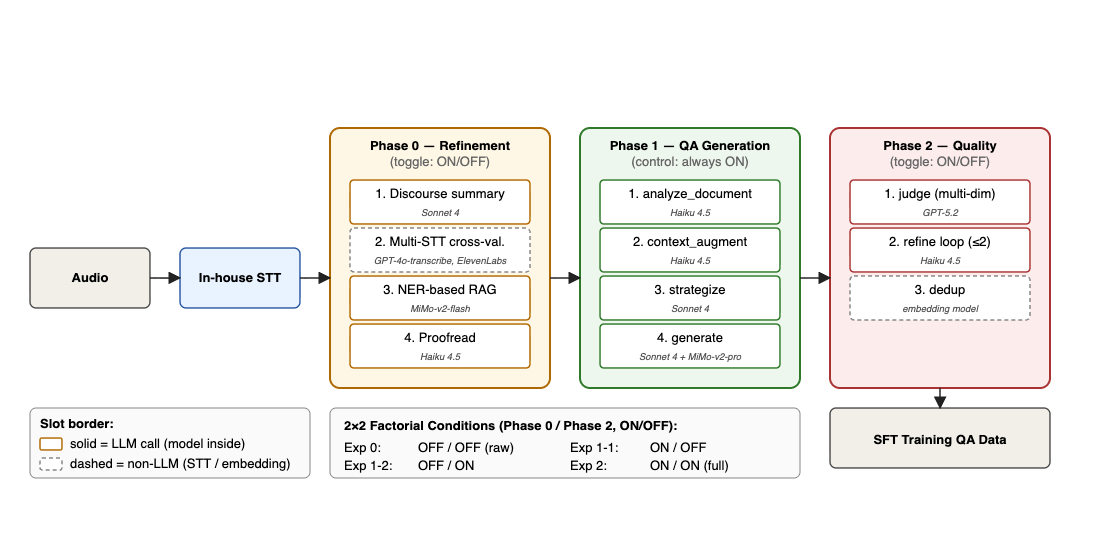}
  \caption{Three-phase speech-to-SFT pipeline. Phase~0 and Phase~2 are independently toggled, yielding the $2\times 2$ factorial; Phase~1 is shared. Phase~0's multi-STT cross-validation re-invokes secondary STT providers on the source audio, normalizing in-house STT differences whenever Phase~0 is active (used in \S\ref{sec:results:whisper}).}
  \label{fig:pipeline}
\end{figure*}

We address both gaps with a $2\times 2$ factorial ablation of a production-ready speech-to-SFT pipeline on Korean medical and finance conference content (Figure~\ref{fig:pipeline}). The two binary stages are \emph{Phase~0} (transcript refinement) and \emph{Phase~2} (quality refinement), with QA generation (Phase~1) shared (Table~\ref{tab:conds}). Each condition produces a distinct SFT dataset from \textbf{40 conference sessions} (19 medical + 21 finance); we train \textbf{9 LLMs (5 families, 2.4B--70B)} with parameter-efficient LoRA, plus a 12-run Full~FT sanity check~\citep{biderman2024lora}. Evaluation spans three core axes, downstream MCQA (KMMLU, KMMLU-Pro, MMLU) after SFT, cell-blind QA-quality scoring by 4 LLM judges~\citep{zheng2023judging} on a 200-QA stratified public sample, and a blind 6-expert human evaluation of QA quality on the same sample, plus two supporting analyses: an STT-engine swap via Whisper-medium~\citep{radford2022whisper} for upstream-STT robustness, and an LLM-difficulty audit by the same 4 LLMs on the same sample under an unknown-admission instruction.

\paragraph{Main contributions.} (i) Under a fixed, standard SFT recipe (\S\ref{sec:setup}), improvements in QA data quality do not transfer uniformly into downstream MCQA gains: 4-judge LLM and 6-expert human evaluations both confirm robust quality improvement ($+0.18$ LLM and $+0.22$ human on a 1--5 scale) while the cross-model mean MCQA gain is not significant; positive transfer concentrates on family$\leftrightarrow$domain-aligned pairs. (ii) We perform a $2\times 2$ factorial ablation of a speech-to-SFT pipeline on a domain-specific case in Korean medical and finance, isolating the independent contributions of transcript refinement and quality refinement. (iii) Two supporting analyses, reported in the appendix, corroborate the pipeline's robustness and the data's difficulty: an STT swap to public Whisper-medium leaves the aggregate downstream effect minimal, at most $1.32$\,pp mean absolute KMMLU difference; and an LLM-difficulty audit shows that even the two frontier LLMs, Opus 4.7 and GPT-5.4, cannot confidently answer $7.8\%$ of the public-sample QA on average. (iv) We release the 200-QA sample plus a Phase-2-filtered variant for sampling sensitivity (Appendix~\ref{app:filtered-sample}), all SFT checkpoints, judge prompts, and analysis code.

\section{Related Work}
\label{sec:related}

\paragraph{Synthetic SFT data from documents.}
A line of work generates SFT data via LLM bootstrapping from seeds or web text: Self-Instruct~\citep{wang2023selfinstruct}, Evol-Instruct~\citep{xu2024wizardlm}, and Cosmopedia~\citep{benallal2024cosmopedia}. Document-to-SFT pipelines such as EntiGraph~\citep{yang2025entigraph} and GLAN~\citep{li2024glan} target structured knowledge synthesis. These methods operate on clean text and do not address upstream challenges that arise when the source is speech, namely transcription noise, named-entity grounding, and discourse structure, which a separate line of speech-to-SFT work has begun to explore.

\paragraph{Speech-to-SFT.}
Several recent works generate SFT-style data from speech, addressing some of these upstream challenges: COSMIC~\citep{pan2024cosmic} synthesizes speech-instruction QA via GPT-3.5 on 450 hours of English audio for speech in-context learning; SIFT-50M~\citep{amazon2025sift} is a 50M-example multilingual speech instruction dataset; LiveCC~\citep{chen2025livecc} interleaves YouTube ASR captions with video frames for streaming video commentary; PodGPT~\citep{li2025podgpt} repurposes medical podcast transcripts but applies continuous pre-training (CPT) rather than SFT. All of these target general capabilities or use CPT rather than text-LLM SFT; none of them perform a per-stage ablation of refinement stages. We close both gaps with a $2\times 2$ factorial ablation on a domain-specific case.

\paragraph{LoRA and LLM-as-judge methodology.}
We use QLoRA~\citep{dettmers2023qlora}, a quantized variant of LoRA~\citep{hu2022lora}, for parameter-efficient SFT; \citet{biderman2024lora} recently showed that LoRA ``learns less and forgets less'' than full fine-tuning, motivating our Full~FT sanity check (\S\ref{sec:results:fullft}). Our QA-data-quality evaluation follows the LLM-as-judge framework~\citep{zheng2023judging} with four cross-provider judges and cell-blind inputs to mitigate within-provider and source-identification bias.

\paragraph{Korean and cross-lingual MCQA benchmarks.}
KMMLU~\citep{son2024kmmlu} provides 35K Korean MCQA across 45 subjects but covers medicine only sparsely and finance via six subjects. KMMLU-Pro~\citep{son2024kmmlu_pro} extends KMMLU with Korean professional-license-exam items covering both medical and finance (Appendix~\ref{app:b}). We additionally use MMLU~\citep{hendrycks2021mmlu} as an English cross-lingual sanity benchmark.

\section{Pipeline and Dataset}
\label{sec:pipeline}

\subsection{Pipeline architecture}
\label{sec:pipeline:arch}

Each phase decomposes into sub-steps with distinct model assignments (Figure~\ref{fig:pipeline}). The two refinement stages target distinct failure modes of naive speech-to-SFT: Phase~0 addresses transcription noise, missing named entities, and broken discourse in raw STT; Phase~2 addresses variability in the faithfulness and depth of generated QA; Phase~1 is held fixed so the ablation isolates refinement rather than generation. \textbf{Phase~0} (Transcript Refinement) cleans sentence-segmented STT output through discourse summary, multi-STT cross-validation against secondary STT providers, NER-based RAG, and a unified proofreading LLM. \textbf{Phase~1} (QA Generation) converts the transcript to QA via analyze + context augment + strategize + generate; each item carries metadata used verbatim by downstream evaluation: \texttt{domain}, \texttt{sub\_domain}, \texttt{difficulty} (\{easy, medium, hard\}), and \texttt{question\_type} (\{factoid, explanatory, procedural, comparative\}). \textbf{Phase~2} (Quality Refinement) applies judge filtering (5-judge ensemble), a refinement loop of at most two rounds, discarding items that still fail, and embedding-based deduplication. The verbatim Phase~2 prompts are released, since Phase~2 is the measurement instrument of our ablation.

\begin{table}[t]
\centering\small
\begin{tabular}{lccl}
\toprule
Condition & Phase 0 & Phase 2 & Label \\
\midrule
Exp~0 & OFF & OFF & no refinement \\
Exp~1-1 & ON & OFF & Phase 0 only \\
Exp~1-2 & OFF & ON & Phase 2 only \\
Exp~2 & ON & ON & full pipeline \\
\bottomrule
\end{tabular}
\caption{$2\times 2$ factorial conditions.}
\label{tab:conds}
\end{table}

\subsection{Dataset statistics}
\label{sec:pipeline:dataset}
The dataset comprises 40 conference sessions drawn from the subset of session data collected through our service for which speakers consented to research and non-commercial use. The released 200-QA public sample is filtered for personally identifiable information (PII). The 40 sessions span 19 medical and 21 finance, primarily Korean with multilingual segments preserved in the source language; medical sub-domains include surgery, cardiology, endocrine, dental, and oncology, and finance sub-domains include VC, banking, fintech, AI policy, and trade. Per-condition QA counts span 2{,}598--2{,}765 ($\sim$6\% range, Table~\ref{tab:counts}), narrow enough not to materially confound factorial comparisons: Phase~0 raises Phase~1 candidate yield, while Phase~2 removes a small fraction per session (per-stage rates in Appendix~\ref{app:stage-acceptance}, Table~\ref{tab:app:stage}). See Appendix~\ref{app:sessions}, Table~\ref{tab:app:sessions} for session-level metadata.

\begin{table}[t]
\centering\small
\begin{tabular}{lrrr}
\toprule
Condition & Medical & Finance & Total \\
\midrule
Exp~0 & 1{,}411 & 1{,}252 & 2{,}663 \\
Exp~1-1 & 1{,}468 & 1{,}297 & 2{,}765 \\
Exp~1-2 & 1{,}402 & 1{,}270 & 2{,}672 \\
Exp~2 & 1{,}376 & 1{,}222 & 2{,}598 \\
\bottomrule
\end{tabular}
\caption{QA counts per condition.}
\label{tab:counts}
\end{table}

\subsection{Sample selection}
\label{sec:pipeline:sampling}
We extract 200 QA (4 conditions $\times$ 2 domains $\times$ 25) for LLM-judge and human QA-quality evaluations (\S\ref{sec:results:judge}) and LLM-difficulty audit (\S\ref{sec:results:llmdiff}). Selection uses sub-domain stratification and random selection within strata. The 200-QA sample is drawn from the training set rather than held out; evaluation in \S\ref{sec:results:judge} and \S\ref{sec:results:llmdiff} does not query the SFT models we train, so this does not introduce evaluation leakage. We additionally construct a parallel 200-QA selection limited to items passing the Phase-2 judges, applied uniformly across all four conditions (Appendix~\ref{app:filtered-sample}), as a sampling-design sensitivity check.

\section{Experimental Setup}
\label{sec:setup}

\paragraph{Models.} Throughout this paper a \emph{cell} denotes one (model $\times$ condition $\times$ domain) unit; the main factorial spans 72 such cells. We evaluate 9 LLMs released after September 2024, spanning 5 families (EXAONE, Gemma, Llama, Phi, Qwen) and 3 size tiers: small (EXAONE 3.5 2.4B, Gemma 3 4B, Llama 3.2 3B, Phi-4 Mini, Qwen 3.5 4B), medium (EXAONE 3.5 7.8B, Qwen 3.5 9B), large (Gemma 3 27B, Llama 3.3 70B). We perform 72 main LoRA runs ($9\!\times\!4\!\times\!2$), 12 Full~FT sanity runs on 3 models (EXAONE 3.5 2.4B, Qwen 3.5 4B, EXAONE 3.5 7.8B) $\times$ Exp~0/2 $\times$ 2 domains (Appendix~\ref{app:k}, Table~\ref{tab:app:fullft}), and an 8-cell Whisper-medium swap grid (W-grid; \{EXAONE 3.5 7.8B, Qwen 3.5 4B\} $\times$ \{medical, finance\} $\times$ \{W-Exp0, W-Exp2\}) for the STT-swap robustness analysis in \S\ref{sec:results:whisper}; W-Exp$k$ substitutes Whisper-medium for the in-house transcript at Exp~$k$.

\paragraph{Training.} LoRA uses rank $r\!=\!16$, scaling $\alpha\!=\!32$ (the $\alpha\!=\!2r$ convention adopted by \citet{biderman2024lora}), dropout 0.05, QLoRA 4-bit, learning rate $2{\times}10^{-4}$ with 3\% warmup, 3 epochs, and effective batch 16. Full~FT uses ZeRO-3. All main LoRA cells run with $n\!=\!2$ seeds; auxiliary experiments use varying seeds (full inventory in Table~\ref{tab:app:seed-inventory}).

\paragraph{Evaluation.}
\label{sec:setup:eval}
Evaluation spans three core axes: (i) cell-blind 4-judge LLM QA-quality scoring, (ii) a blind 6-expert human evaluation of QA quality, and (iii) downstream MCQA accuracy after SFT. Two supporting analyses, (iv) an STT-swap robustness check via Whisper-medium and (v) an LLM-difficulty audit, are reported mainly in the appendix. For (i) and (v) we use four cross-provider LLMs (frontier: Opus 4.7, GPT-5.4; production: Sonnet 4.6, GPT-4o), serving as judges in (i) and answering the public sample in (v). (i) \textbf{LLM QA quality}: the four LLMs score the 200-QA public sample under a 1--5 Likert rubric on five external-quality dimensions (faithfulness, domain\_accuracy, question\_quality, answer\_depth, coherence)~\citep{zheng2023judging} with cell-blind inputs, withholding condition and source identifiers. 4-judge mean and per-judge breakdown in Appendix~\ref{app:judge}. (ii) \textbf{Human QA quality}: three medical and three finance domain experts independently score the same 100-QA per-domain sample that the LLM judges scored, under the identical 5-dim Likert rubric and cell-blind protocol. Per-domain agreement, per-rater results, and profiles in Appendix~\ref{app:human}. (iii) \textbf{MCQA}: KMMLU (45 Korean subjects), evaluated overall and on a training-domain-aligned subset (Appendix~\ref{app:b}); KMMLU-Pro~\citep{son2024kmmlu_pro} with license-aligned filtering; MMLU (57 subjects) evaluated overall, and training-domain-aligned in Full~FT (\S\ref{sec:results:fullft}). (iv) \textbf{STT-swap robustness (W-grid)}: we regenerate QA from Whisper-medium~\citep{radford2022whisper} on the matched EXAONE 3.5 7.8B + Qwen 3.5 4B $\times$ medical + finance cells. (v) \textbf{LLM-difficulty audit}: the same four LLMs answer the public sample under an unknown-admission instruction.

\paragraph{Statistics.} Let $\Delta_{2-0}\!\equiv\!\text{Exp 2}\!-\!\text{Exp 0}$ denote the full-pipeline effect. We report two factorial decompositions: \emph{QA data quality} (\S\ref{sec:results:judge}) uses single-stage effects $P_0^{s}\!=\!\text{Exp 1-1}\!-\!\text{Exp 0}$ and $P_2^{s}\!=\!\text{Exp 1-2}\!-\!\text{Exp 0}$ (identity $P_0^{s}\!+\!P_2^{s}\!+\!P_0\!\times\!P_2\!=\!\Delta_{2-0}$, with $P_0\!\times\!P_2\!=\!\text{Exp 2}\!-\!\text{Exp 1-1}\!-\!\text{Exp 1-2}\!+\!\text{Exp 0}$ shared); \emph{MCQA} (\S\ref{sec:results:factorial}) uses standard $2{\times}2$ marginal main effects $P_0^{a}, P_2^{a}$ (averaged over the other factor; identity $P_0^{a}\!+\!P_2^{a}\!=\!\Delta_{2-0}$). The single-stage decomposition surfaces the interaction; the marginal decomposition averages over the other factor. $\Delta_{2-0}$ is identical under both. Per-sample bootstrap 95\% CIs ($100{,}000$ resamples) in Appendix~\ref{app:bootstrap}, Table~\ref{tab:app:bootstrap}. A mixed-effects regression with cell-level random intercepts (\S\ref{sec:results:factorial}) separates between-cell and seed-level variance.

\section{Results}
\label{sec:results}

Following the pipeline order, we first report the three core evaluations: QA data quality by LLM judges and by human experts at the generation stage (\S\ref{sec:results:judge}), then downstream MCQA after SFT (\S\ref{sec:results:factorial}). Two supporting analyses, an STT-swap robustness check (\S\ref{sec:results:whisper}) and an LLM-difficulty audit (\S\ref{sec:results:llmdiff}), are summarized here with full detail in the appendix.

\begin{table*}[t]
\centering\small
\begin{tabular}{lrrrrrrrr}
\toprule
Source & Exp~0 & Exp~1-1 & Exp~1-2 & Exp~2 & $P_0^{s}$ & $P_2^{s}$ & $P_0{\times}P_2$ & $\Delta_{2-0}$ \\
\midrule
LLM (4-judge, Finance)       & 3.76 & 3.84 & 3.76 & 3.93 & $+0.08$ & $+0.00$ & $+0.09$ & $+0.17$ \\
LLM (4-judge, Medical)       & 3.85 & 3.95 & 3.92 & 4.05 & $+0.10$ & $+0.06$ & $+0.03$ & $+0.20$ \\
LLM (2 frontier, Finance)    & 3.36 & 3.48 & 3.40 & 3.57 & $+0.11$ & $+0.04$ & $+0.06$ & $+0.20$ \\
LLM (2 frontier, Medical)    & 3.45 & 3.52 & 3.49 & 3.72 & $+0.07$ & $+0.04$ & $+0.16$ & $\mathbf{+0.26}$ \\
\midrule
Human (Finance, 3 raters)    & 3.00 & 3.20 & 3.13 & 3.29 & $+0.21$ & $+0.14$ & $-0.05$ & $+0.29$ \\
Human (Medical, 3 raters)    & 3.72 & 3.73 & 3.78 & 3.87 & $+0.01$ & $+0.06$ & $+0.08$ & $+0.15$ \\
\bottomrule
\end{tabular}
\caption{QA data quality on the 200-QA public sample (1--5 scale; 5-dim mean). All six rows show positive $\Delta_{2-0}$; cross-source domain pattern is consistent. Averaging Finance and Medical rows recovers the overall values: 4-judge $\Delta_{2-0}\!=\!+0.18$, 2 frontier $\Delta_{2-0}\!=\!+0.23$, human (6-rater) $\Delta_{2-0}\!=\!+0.22$. Per-dimension LLM-judge breakdown in Appendix~\ref{app:judge_per_dim}, Table~\ref{tab:app:judge_per_dim}; per-judge LLM breakdown in Appendix~\ref{app:judge}, Table~\ref{tab:app:judge}; per-rater human results in Appendix~\ref{app:human}, Table~\ref{tab:app:human_per_rater}. Cells rounded independently; $P_0^{s}\!+\!P_2^{s}\!+\!P_0\!\times\!P_2$ may differ from $\Delta_{2-0}$ by $\pm 0.01$.}
\label{tab:judge}
\end{table*}

\subsection{QA data quality}
\label{sec:results:judge}

Table~\ref{tab:judge} reports the overall score (5-dim mean) by source and condition; per-dim LLM-judge breakdown is in Appendix~\ref{app:judge_per_dim}.

\paragraph{Full-pipeline effect and factor decomposition.}
The 4-judge mean progresses Exp~0 $3.81 \rightarrow$ Exp~1-1 $3.89 \rightarrow$ Exp~1-2 $3.84 \rightarrow$ Exp~2 $3.99$, giving $\Delta_{2-0}\!=\!+0.18$ (95\% CI $[+0.06, +0.32]$) on the 5\,pt scale (Figure~\ref{fig:judgemcqa}a, Table~\ref{tab:judge}); the frontier subset lifts this to $\mathbf{+0.23}$ (95\% CI $[+0.03, +0.44]$). Phase~0 is the stronger single-stage contributor ($P_0^{s}\!=\!+0.09$ vs $P_2^{s}\!=\!+0.03$), favored by three of four judges, Sonnet 4.6, Opus 4.7, and GPT-5.4; GPT-4o is the exception ($P_2^{s}\!>\!P_0^{s}$ by $+0.03$). All five dimensions improve from Exp~0 to Exp~2 (Appendix~\ref{app:judge_per_dim}); per-judge breakdown in Appendix~\ref{app:judge}, Table~\ref{tab:app:judge}.

\paragraph{Phase-2 mechanism: composition shift and within-type quality.}
Phase~2 shifts question-type composition toward more explanatory questions (Appendix~\ref{app:sample}, Table~\ref{tab:typecomp}): Exp~2 is explanatory-dominant ($+10$\,pp vs.\ Exp~0), while Phase~0 alone (Exp~1-1) instead lifts factoid. Type-stratified analysis (Appendix~\ref{app:sample}) shows within-type gains positive on all 4 types; re-weighting Exp~2 to Exp~0's composition yields type-controlled $\Delta_{2-0}\!\approx\!+0.16$, so within-type improvement accounts for $\sim\!86\%$ of $\Delta_{2-0}$ and composition shift $\sim\!14\%$ (per-type $n\!=\!5$--$20$, Table~\ref{tab:app:typestrat}). On the Phase-2-filtered sensitivity sample, the 4-judge $\Delta_{2-0}$ matches the primary $+0.18$ within $0.01$, so the Phase~2 effect is robust to the sampling design.

\paragraph{Human evaluation results.}
All six raters, including all three finance raters, independently show a positive $\Delta_{2-0}$, with a cross-rater pooled $\Delta_{2-0}$ of $+0.220$, or $+0.370\sigma$ after z-normalization, in close agreement with the LLM-judge $\Delta_{2-0}\!=\!+0.18$. Inter-rater agreement is ICC(2,3)$\!=\!0.69$ on medical and $0.34$ on finance, with Krippendorff $\alpha\!=\!0.42$ and $0.03$; the lower finance figure reflects absolute-scale spread across raters rather than directional disagreement, as the per-rater $\Delta_{2-0}$ direction is unanimous. Per-rater scores, pooled means, and inter-rater statistics in Appendix~\ref{app:human}.

\subsection{Factorial main effects on MCQA}
\label{sec:results:factorial}

\begin{table}[!t]
\centering\small
\setlength{\tabcolsep}{4pt}
\begin{tabular}{lrrrr}
\toprule
Metric & $P_0^{a}$ & $P_2^{a}$ & $P_0{\times}P_2$ & $\Delta_{2-0}$ \\
\midrule
MMLU & $-0.07$ & $+0.10$ & $+0.12$ & $+0.04$ \\
KMMLU & $+0.45$ & $+0.05$ & $+0.28$ & $+0.50$ \\
KMMLU\,(med) & $+0.21$ & $-0.03$ & $-0.04$ & $+0.18$ \\
KMMLU\,(fin) & $+0.09$ & $+0.35$ & $+0.51$ & $+0.45$ \\
KMMLU-Pro\,(med) & $-0.17$ & $+0.26$ & $-0.36$ & $+0.09$ \\
KMMLU-Pro\,(fin) & $-0.08$ & $+0.29$ & $-0.29$ & $+0.21$ \\
\bottomrule
\end{tabular}
\caption{Cross-model factorial effects on MCQA (pp). \texttt{(med)/(fin)} rows: training-domain-aligned subsets (KMMLU by subject, Appendix~\ref{app:b}; KMMLU-Pro by license, Appendix~\ref{app:kmmlu_pro}, Tables~\ref{tab:app:kmmlu_pro} and~\ref{tab:app:kmmlu_pro_large}). Aggregates use unrounded cell means; within-table $P_0^{a}\!+\!P_2^{a}$ may differ from $\Delta_{2-0}$ by $\pm 0.01$\,pp.}
\label{tab:factorial}
\end{table}

\begin{figure}[!t]
\centering
\includegraphics[width=\columnwidth]{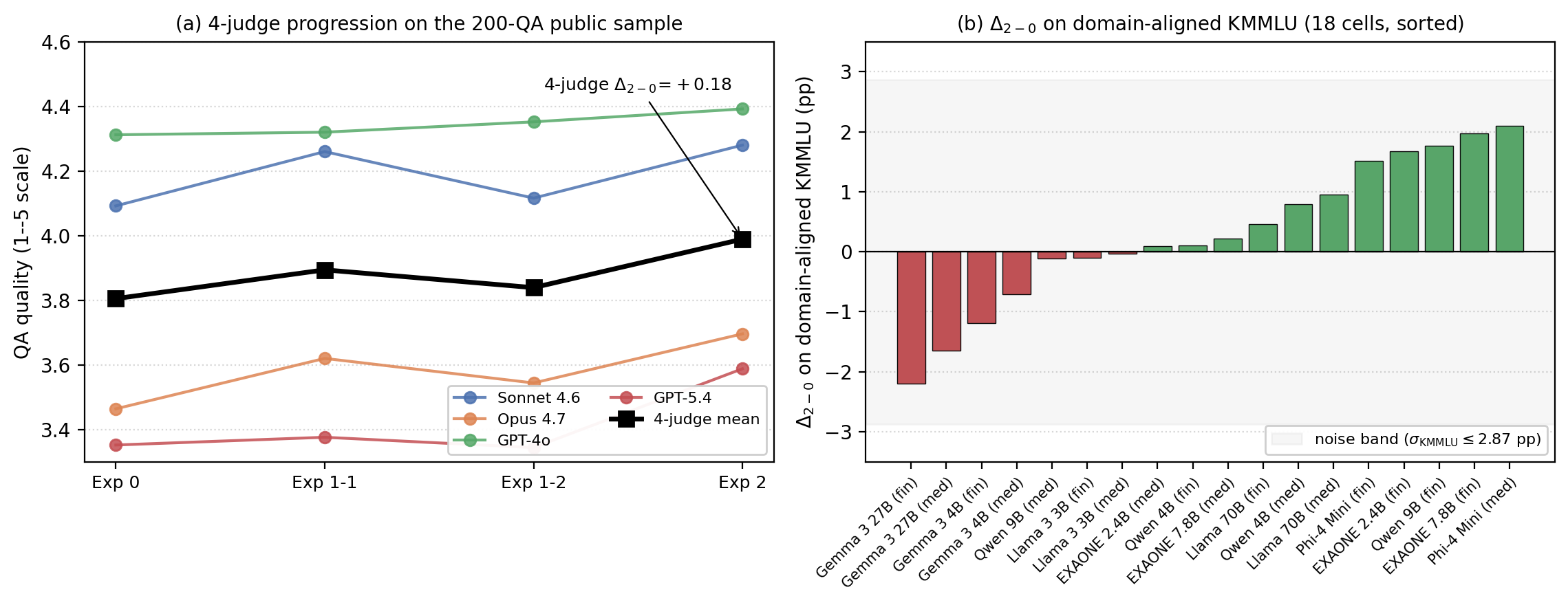}
\caption{(a) 4-judge progression on the 200-QA public sample (per-judge means; 4-judge mean overlaid, $\Delta_{2-0}\!=\!+0.18$). (b) $\Delta_{2-0}$ on the domain-aligned KMMLU subset, 18 cells sorted; shaded band is the per-cell noise bound $\sigma_{\text{KMMLU}}\!\le\!2.87$\,pp.}
\label{fig:judgemcqa}
\end{figure}

Table~\ref{tab:factorial} reports the factorial decomposition into Phase~0 and Phase~2 main effects ($P_0^{a}$, $P_2^{a}$) and their interaction across six MCQA aggregates from the $9 \times 4 \times 2$ design.

\paragraph{Phase-specific dominance.} The two MCQA benchmark families respond to different pipeline stages: Phase~0 lifts KMMLU overall ($P_0^{a}\!=\!+0.45$ vs $P_2^{a}\!=\!+0.05$), while Phase~2 lifts KMMLU-Pro ($P_2^{a}\!=\!+0.26$/$+0.29$ vs $P_0^{a}\!=\!-0.17$/$-0.08$ on medical/finance). On the KMMLU domain-aligned subsets, Phase~0 leads on medical and Phase~2 on finance; MMLU is compressed across all effects.

\paragraph{Full-pipeline effect $\Delta_{2-0}$.} All six aggregates show positive $\Delta_{2-0}$ ($+0.04$ to $+0.50$\,pp), with per-cell signs model-family-conditional. $P_0\!\times\!P_2$ varies in sign and magnitude across aggregates, so cross-model aggregates reflect family-conditional rather than uniform effects.

\paragraph{Per-cell pattern.} Figure~\ref{fig:judgemcqa}b shows positive cells on the domain-aligned KMMLU subset concentrate on finance-trained EXAONE 3.5 7.8B ($+1.98$\,pp) and Qwen 3.5 9B ($+1.71$\,pp), and medical-trained Phi-4 Mini ($+2.10$\,pp); negative cells cluster on both Gemma 3 sizes (27B: $-1.46$ med, $-2.11$ fin). Per-cell raw scores are in Appendix~\ref{app:all72}, Table~\ref{tab:app:all72}. Stage-wise SFT transfer is therefore conditional on family$\leftrightarrow$domain alignment (\S\ref{sec:discussion}).

\paragraph{Multi-seed stability and bootstrap CI.}
Per-cell seed $\sigma_{\text{KMMLU}}\!\le\!2.87$\,pp worst-case; most cells $<\!1$\,pp. The strongest positive cell ($\Delta_{2-0}\!=\!+1.98$\,pp at $\sigma\!=\!1.41$\,pp) is a ${\sim}1.4\sigma$ effect. A paired bootstrap over the 18 (model$\times$domain) cells gives cross-cell mean $\Delta_{2-0}\!=\!+0.31$\,pp with 95\% CI $[-0.24, +0.86]$ (paired $t$ $p\!=\!0.28$), consistent with a family-conditional signal rather than a uniform improvement. A mixed-effects analysis attributes $99.4\%$ of the total variance to between-cell (model$\times$domain) heterogeneity rather than seed noise (Appendix~\ref{app:mixedeffects}), so per-cell effects reflect cell-level signal rather than seed-level instability.

\subsection{STT-swap robustness}
\label{sec:results:whisper}

We regenerate QA from public Whisper-medium at Exp~0 and Exp~2 and re-train the matched EXAONE 3.5 7.8B and Qwen 3.5 4B cells across medical and finance; conclusions at both evaluation layers are unchanged. QA data quality holds under the same 4-judge protocol, with across-cell mean $|\Delta|\!\le\!0.21$ (Table~\ref{tab:lt_vs_w}), and downstream accuracy stays within the per-cell noise band, at most $1.32$\,pp mean absolute deviation on KMMLU and $0.39$\,pp on MMLU. Upstream WER and per-cell detail are in Appendices~\ref{app:wer} and~\ref{app:multiseed}.

\subsection{LLM-difficulty audit}
\label{sec:results:llmdiff}

Four LLMs answer the 200-QA public sample under an unknown-admission instruction. Under a cell-blind 2-judge strict re-classification, with judge agreement Cohen's $\kappa\!=\!0.903$, the two frontier models cannot confidently answer $7.8\%$ of the sample (Table~\ref{tab:llmdiff}), a low-cost triage signal for the session-grounded corpus. Per-model, per-condition, and keyword-heuristic detail are in Appendix~\ref{app:keyword}.

\subsection{Full FT sanity check}
\label{sec:results:fullft}
A 12-run Full~FT sanity (3 models, Exp~0/2, 2 domains; Appendix~\ref{app:k}, Table~\ref{tab:app:fullft}) corroborates LoRA fidelity at medium scale: per-cell Exp~2$-$base $r\!=\!0.841$ ($p\!<\!0.001$) and Exp~2$-$Exp~0 $r\!=\!0.648$ ($p\!\approx\!0.022$, $n\!=\!12$); sign agreement is $11/12$ on Exp~2$-$base and $7/12$ on Exp~2$-$Exp~0, with 4 of 5 flips concentrated on the two smallest models, consistent with LoRA instability at high rank-ratio on small scale~\citep{biderman2024lora}.

\section{Discussion}
\label{sec:discussion}

\paragraph{Cross-instrument convergence.} The quality improvement direction is consistent across the four cross-provider LLM judges and the six human experts, with per-judge $\Delta_{2-0}$ spreading from $+0.08$ to $+0.24$ (Appendix~\ref{app:judge}). The agreement strengthens confidence in the quality finding beyond single-instrument bias.

\paragraph{QA quality gains do not transfer uniformly downstream.} Under our standard SFT recipe, 4-judge LLM and 6-expert human evaluations both confirm robust QA quality improvement ($\Delta_{2-0}\!=\!+0.18$ and $+0.22$), yet the cross-cell mean MCQA $\Delta_{2-0}$ is not significant; positive transfer concentrates on family$\leftrightarrow$domain-aligned cells, so positive and negative cells largely cancel in cross-model averages. One candidate account of this pattern is a format mismatch, since Phase~2 shifts SFT data toward explanatory items while MCQA probes factoid recall; we frame this as a candidate account rather than a confirmed mechanism. The SFT recipe is held fixed across all cells by design, to isolate the contribution of each pipeline stage; how downstream transfer depends on the recipe is a separate, well-studied axis that we leave to future work.

\paragraph{Hardness of the data.} The two frontier LLMs admit unknown on $7.8\%$ of QA (Table~\ref{tab:llmdiff}), reflecting the session-grounded, institution-specific character of the source audio. This bound offers a low-cost triage tool for new corpora and quantifies the value of session-grounded SFT data over open-web augmentation: low admit-rate corpora likely overlap with web pretraining and offer limited SFT differentiation.

\paragraph{Industry implications.} Cross-cell mean KMMLU $\Delta_{2-0}$ is not significant and dominance is family-conditional, so per-cell validation is needed before scaling to new model families or fixed deployments. Phase~0's multi-STT cross-validation lets a public STT engine substitute for in-house STT at the aggregate level. Phase~2 best fits QA-format-aligned downstream tasks. Per-phase token costs are reported in Appendix~\ref{app:cost}.

\label{body:end}

\section*{Limitations}
\label{sec:limits}

\paragraph{Seed coverage.} All 72 main LoRA cells are validated with $n\!=\!2$ seeds ($\sigma_{\text{KMMLU}}\!\le\!2.87$\,pp, $\sigma_{\text{MMLU}}\!\le\!0.66$\,pp worst-case across cells; per-seed evaluation outputs in the companion repository; cell means in Appendix~\ref{app:all72}, Table~\ref{tab:app:all72}). Multi-seed reporting is uncommon in the LoRA SFT literature, where a single fixed seed is the dominant practice~\citep{biderman2024lora}; $n\!=\!2$ per cell exceeds this norm. We further rely on cross-model aggregation (18 model$\times$domain pairs) for robustness. The 12 Full~FT sanity cells use seed 42 only. The W-grid uses $n\!=\!2$ seeds. Full per-cell seed inventory in Table~\ref{tab:app:seed-inventory}.

\paragraph{Catastrophic forgetting.} We additionally evaluated 8 SFT cells (12 evaluation runs: 4 multi-seed + 4 legacy single-seed Whisper STT analysis cells) on KoBEST, a Korean general-purpose benchmark with 5 subtasks (Appendix~\ref{app:kobest}, Table~\ref{tab:app:kobest}). Average KoBEST accuracy is stable across conditions and close to base, with only a mild $\sim\!2$\,pp drop on Llama 3.3 70B. The small MCQA effects in \S\ref{sec:results:factorial} are not explained by catastrophic forgetting of general Korean capability.

\paragraph{Sample-bound external claims.} The LLM-difficulty audit and LLM-judge QA data quality are measured on a 200-QA public sample, not the full training set; stratified random selection is used but a small sample-bias risk remains.

\paragraph{Cross-judge variance.} Per-judge $\Delta_{2-0}$ spreads from $+0.08$ to $+0.24$ (Appendix~\ref{app:judge}, Table~\ref{tab:app:judge}); 3 of 4 judges favor Phase~0 on the single-stage decomposition (\S\ref{sec:results:judge}).

\paragraph{Partial formal statistical tests.} We did not run a multiple-comparison correction across all 72 individual cells. For the subset of cells with per-sample evaluation data preserved, 8 cells covering high-signal multi-seed and Whisper STT analysis runs (Appendix~\ref{app:bootstrap}, Table~\ref{tab:app:bootstrap}), we report bootstrap 95\% CIs (100{,}000 resamples): typical width $\pm0.5$\,pp on KMMLU and $\pm0.8$\,pp on MMLU; seed variance ($n\!=\!2$ per cell, worst-case $\sigma\!=\!2.87$\,pp) dominates per-sample bootstrap variance. For the 18 cross-model (model$\times$domain) cells we additionally report a paired bootstrap CI and paired $t$ test on the cross-cell mean $\Delta_{2-0}$ in \S\ref{sec:results:factorial}: the cross-cell average domain-aligned KMMLU effect ($+0.31$\,pp, 95\% CI $[-0.24,+0.86]$) is not statistically distinguishable from zero. This null is informative: the $+0.86$\,pp CI upper bound excludes large uniform effects rather than reflecting limited power. Appendix~\ref{app:mixedeffects} reports a mixed-effects model with the Exp~0$\to$Exp~2 toggle as a fixed effect and (model, domain) cell as a random intercept ($\hat\beta\!=\!+0.317$\,pp, 95\% CI $[-0.06, +0.69]$, $p\!=\!0.096$), consistent with this bound but estimating only the average effect. Allowing this effect to vary across model families, via a family$\times$toggle interaction or family-level random slopes, would formally quantify the family-conditional pattern (\S\ref{sec:results:factorial}) and is left as future work.

\paragraph{SFT recipe scope.} The $2\times 2$ ablation varies only the data pipeline (Phase~0, Phase~2); the SFT recipe (parameter-efficient LoRA, lr $2\!\times\!10^{-4}$, 3 epochs, QLoRA 4-bit) is held fixed across all 72 main cells, with the 12-run Full~FT sanity check as the only deviation. Sensitivity of the downstream transfer signal to alternative SFT recipes, such as longer training, different objectives, larger data, or curricula, lies outside this data-side study and is appropriate for downstream training-pipeline studies to characterize.

\paragraph{Downstream evaluation scope.} Format-aligned downstream evaluations such as open-ended QA or RAG-QA may complement the MCQA endpoint; broader endpoint coverage is left as future work.

\paragraph{Phase-0 sub-ablation.} Phase~0's internal sub-steps, summary, multi-STT, NER-based RAG, and proofread, are evaluated as a single block; isolating each is left as future work.

\paragraph{STT coverage.} We selected Whisper-medium to evaluate pipeline resilience under high-noise conditions, maximizing the upstream STT divergence from our in-house STT; Whisper large-v3, ElevenLabs, Soniox, and similar production STT engines remain future work. The W-grid focuses on the medium tier, EXAONE 3.5 7.8B and Qwen 3.5 4B, representative of typical production-scale deployment, across medical + finance at both W-Exp0 and W-Exp2; broader tier coverage is left as future work.

\paragraph{LLM unknown-admission detection.} The 2-judge binary protocol (\S\ref{sec:results:llmdiff}, Table~\ref{tab:llmdiff}) shows large per-provider spread, with GPT-4o concentrating explicit unknown-admissions.

\paragraph{Per-condition QA-count variation.} Condition-level QA counts vary by up to 167 items (Table~\ref{tab:counts}, ${\sim}6\%$ spread). This drift is an intrinsic outcome of the phase mechanisms themselves rather than a confound; subsampling to equalize counts would itself bias the comparison by removing items the pipeline selectively retains. Question-type composition shifts are discussed in \S\ref{sec:results:judge} as Phase-2 structural downstream effects rather than confounds.

\section*{Ethics Statement}

\paragraph{Data collection and consent.} The 40 conference sessions used in this work are a subset of Korean-language session data collected through an AI-mediated interpretation service, spanning academic conferences, professional seminars, and industry forums in the medical and finance sectors (2024--2026); speakers consented to research and non-commercial use. The source audio, full transcripts, and the complete 10{,}698-QA generated corpus are not released, in accordance with the consent scope. The original audio remains subject to speaker-consent agreements and is not redistributable.

\paragraph{Personally identifiable information.} The released 200-QA public sample and the Phase-2-filtered variant have personally identifiable information replaced with neutral placeholders (e.g., \texttt{[Brokerage A]}, \texttt{[Overseas University Medical Center]}). Generic external organization names that appear as domain context and are not PII are retained as discourse anchors; these are not authors' affiliations.

\paragraph{Human evaluator recruitment.} The six human evaluators used in our blind expert evaluation, three medical and three finance, were recruited externally through referral from in-house contacts. No evaluator had an employment relationship with the authors or their institution, nor participated in pipeline design or model selection. Evaluators received compensation for their participation; only anonymized professional profiles comprising role, sub-domain, and years-of-experience range are disclosed. Evaluation protocol details are described in \S\ref{sec:setup:eval} and Appendix~\ref{app:human}.

\paragraph{Released artifacts and constraints.} The companion repository \url{https://github.com/flitto/speech-to-sft-ablation-paper} discloses all pipeline LLM identifiers, training scripts, configs, evaluation rubrics, the Phase~2 verbatim prompts (judge, refine, dedup), and the 200-QA primary sample plus the Phase-2-filtered variant. SFT checkpoints (172 runs) are released at \url{https://huggingface.co/collections/Flitto/expert-qa-pipeline-emnlp-2026-industry}. The full pipeline implementation is available upon reasonable request. Downstream use of the released artifacts should respect the consent scope (research and non-commercial use).

\paragraph{Code of ethics.} This work adheres to the ACL Code of Ethics. Human-subjects involvement is limited to the consent-given source dataset and the blind 6-expert evaluation described in \S\ref{sec:setup:eval} and Appendix~\ref{app:human}, conducted under voluntary participation with anonymized professional profile disclosure only.

\bibliography{references}

\appendix

\section{Per-phase token costs}
\label{app:cost}

Per-QA token-cost estimates (USD; OpenRouter unified pricing, snapshot 2026-05), measured on a representative log sub-sample:
\begin{itemize}
\item Phase~0: 0.011--0.014 (multi-STT + NER-based RAG + proofread)
\item Phase~1: 0.040--0.050 (analyze + augment + strategize + generate)
\item Phase~2: 0.009--0.013 (judge + refine $\leq 2$ + dedup)
\end{itemize}

\section{Seed inventory}
\label{app:seed-inventory}

Table~\ref{tab:app:seed-inventory} lists the seed configuration used in every experiment reported in this paper.

\begin{table}[!p]
\centering\footnotesize
\setlength{\tabcolsep}{4pt}
\begin{tabular}{lrl}
\toprule
Experiment & Cells & Seeds \\
\midrule
Main LoRA SFT & 72 & $\{1, 2\}$ ($n{=}2$) \\
MCQA eval (KMMLU/MMLU) & 72 & $\{1, 2\}$ ($n{=}2$) \\
KMMLU-Pro eval & 72 & $\{1, 2\}$ ($n{=}2$) \\
Full FT sanity & 12 & $\{42\}$ ($n{=}1$) \\
W-Exp0 SFT & 4 & $\{1, 2\}$ ($n{=}2$) \\
W-Exp2 SFT & 4 & $\{1, 2\}$ ($n{=}2$) \\
Sample selection (200 QA) & 1 & $\{42\}$ ($n{=}1$) \\
\bottomrule
\end{tabular}
\caption{Seed inventory across experiments. Bootstrap CI and KoBEST evaluations cover a subset of high-signal cells (Appendices~\ref{app:bootstrap}, \ref{app:kobest}).}
\label{tab:app:seed-inventory}
\end{table}

\section{KMMLU and KMMLU-Pro domain mapping}
\label{app:b}

KMMLU 45 subjects $\rightarrow$ domain proxies used in \S\ref{sec:results:factorial}:
\begin{itemize}
\item \textbf{KMMLU-med} (medical-relevant, 4 subjects): \texttt{health}, \texttt{biology}, \texttt{chemistry}, \texttt{psychology}.
\item \textbf{KMMLU-fin} (finance-relevant, 6 subjects): \texttt{economics}, \texttt{accounting}, \texttt{taxation}, \texttt{management}, \texttt{marketing}, \texttt{real\_estate}.
\end{itemize}

MMLU 57 subjects $\rightarrow$ domain proxies used in the Full~FT sanity (\S\ref{sec:results:fullft}, Appendix~\ref{app:k}):
\begin{itemize}
\item \textbf{MMLU-med} (8 subjects): \texttt{clinical\_knowledge}, \texttt{medical\_genetics}, \texttt{anatomy}, \texttt{college\_medicine}, \texttt{professional\_medicine}, \texttt{college\_biology}, \texttt{nutrition}, \texttt{virology}.
\item \textbf{MMLU-fin} (7 subjects): \texttt{econometrics}, \texttt{high\_school\_macroeconomics}, \texttt{high\_school\_microeconomics}, \texttt{professional\_accounting}, \texttt{business\_ethics}, \texttt{management}, \texttt{marketing}.
\end{itemize}

KMMLU-Pro license categories $\rightarrow$ domain tasks (filtered by \texttt{license\_name} via the task YAML's \texttt{process\_docs} hook):
\begin{itemize}
\item \textbf{kmmlu\_pro\_medical} (1{,}205 items): physician, dentist, pharmacist, Korean medicine doctor, Korean pharmacist.
\item \textbf{kmmlu\_pro\_finance} (801 items): certified public accountant, tax accountant, customs broker, certified appraiser.
\end{itemize}
We enforce domain matching at evaluation time: medical-trained SFT cells are scored only on \texttt{kmmlu\_pro\_medical}, finance-trained cells only on \texttt{kmmlu\_pro\_finance}. Per-subject results are released in the companion repository.

\section{Session-level metadata}
\label{app:sessions}

Table~\ref{tab:app:sessions} lists the 40 conference sessions used in the main experiment (19 medical + 21 finance). ROW = transcript line count (sentence-segmented STT output, proxy for session length).

\begin{table}[h]
\centering\footnotesize
\setlength{\tabcolsep}{4pt}
\begin{tabular}{rlr}
\toprule
\# & sub-domain & ROW \\
\midrule
\multicolumn{3}{l}{\textit{Medical (19 sessions)}} \\
\midrule
1  & dental/implant            & 659 \\
2  & dental/AI implant         & 789 \\
3  & dental/PDRN               & 671 \\
4  & dental/implant            & 460 \\
5  & other/critical care       & 419 \\
6  & nursing/trauma            & 484 \\
7  & psychiatry/pain           & 466 \\
8  & surgery/spinal cord       & 477 \\
9  & pediatric/geriatric       & 601 \\
10 & internal/diabetes         & 486 \\
11 & internal/cardiology       & 370 \\
12 & oncology/breast           & 549 \\
13 & oncology/breast tx        & 466 \\
14 & pediatric/geriatric trauma& 616 \\
15 & pharmacy/anticoagulation  & 189 \\
16 & surgery/robotic           & 577 \\
17 & pharmacy/AI Korean med    & 615 \\
18 & psychiatry/therapy        & 221 \\
19 & surgery/plastic           & 781 \\
\midrule
\multicolumn{3}{l}{\textit{Finance (21 sessions)}} \\
\midrule
1  & macro/finance forum       & 663 \\
2  & banking/forum             & 444 \\
3  & VC/banking                & 408 \\
4  & banking/hegemony forum    & 564 \\
5  & accounting/tax            & 203 \\
6  & insurance                 & 373 \\
7  & macro/5-year econ.\ plan  & 409 \\
8  & insurance/welfare lecture & 310 \\
9  & VC/global expansion       & 268 \\
10 & VC/Paris summit           & 733 \\
11 & fintech/fintech week      & 430 \\
12 & securities/Korea econ.    & 697 \\
13 & securities/stock invest.  & 467 \\
14 & crypto/blockchain         & 367 \\
15 & VC/AI solution            & 460 \\
16 & VC/startup conf.          & 723 \\
17 & banking/Japan econ.       & 689 \\
18 & banking/social econ.      & 605 \\
19 & VC/quantum                & 191 \\
20 & ESG/green finance         & 678 \\
21 & fintech/AI summit         & 586 \\
\bottomrule
\end{tabular}
\caption{40 conference sessions used in the main experiment. Columns: index within domain, sub-domain, ROW (transcript-line count).}
\label{tab:app:sessions}
\end{table}

\section{Phase~2 stage acceptance}
\label{app:stage-acceptance}

The per-condition QA-count spread in Table~\ref{tab:counts} arises from the Phase~2 pipeline applied to Phase~1 candidates. Table~\ref{tab:app:stage} reports per-stage acceptance rates on the conditions where Phase~2 is active (Exp~1-2 and Exp~2). Coverage is partial ($n\!=\!6$ Exp~1-2 sessions and $n\!=\!22$ Exp~2 sessions) but the per-stage rates are stable across both conditions.

\begin{table}[h]
\centering\footnotesize
\setlength{\tabcolsep}{4pt}
\begin{tabular}{lrr}
\toprule
Stage & Exp~1-2 ($n\!=\!6$) & Exp~2 ($n\!=\!22$) \\
\midrule
Phase 1 candidates (mean) & 60.0 & 63.1 \\
Judge \emph{pass} rate & 87.8\% & 82.6\% \\
Refine recovery rate & 82.6\% & 78.7\% \\
Dedup duplicates / session & 1.0 & 1.7 \\
Mean exported / session & 56.8 & 58.8 \\
Net Phase~2 retention & 94.7\% & 93.2\% \\
\bottomrule
\end{tabular}
\caption{Phase~2 stage acceptance rates from pipeline logs. \emph{Refine recovery rate} is the fraction of judge-failed items salvaged by the refinement loop (remaining items are discarded). \emph{Net retention} = mean exported / mean Phase 1 candidates. Exp~0 and Exp~1-1 do not run Phase~2 and are reported in Table~\ref{tab:counts} as Phase~1 output only.}
\label{tab:app:stage}
\end{table}

The rate pattern explains the Table~\ref{tab:counts} ordering: Phase~0 raises Phase~1 candidate yield slightly (cleaner transcripts surface more answerable sub-topics), while Phase~2 net-removes $\sim$5--7\% per session ($5.3\%$ Exp~1-2, $6.8\%$ Exp~2) via judge filter + refine discard + dedup. The combined effect keeps per-condition training-set sizes within a $\sim$6\% band.

\section{Full~FT sanity per-cell results}
\label{app:k}

Stage-wise delta $\Delta_{\text{stage}}=\text{Exp}_2-\text{Exp}_0$ on the training-domain-aligned subsets of KMMLU and MMLU across 12 cells (3 models $\times$ 2 domains $\times$ \{KMMLU, MMLU\}) is reported in Table~\ref{tab:app:fullft}. To enable paired comparison with the single-seed Full~FT runs (seed 42 only), the LoRA $\Delta$ column in this table uses the seed-42 LoRA run rather than the $n\!=\!2$ multi-seed mean reported in Appendix~\ref{app:all72}.

\begin{table}[h]
\centering\small
\begin{tabular}{llrrl}
\toprule
Model & Metric & LoRA $\Delta$ & FullFT $\Delta$ & Sign \\
\midrule
exaone3.5-2.4b & KMMLU-med & $-2.44$ & $-0.14$ & $\checkmark$ \\
exaone3.5-2.4b & MMLU-med  & $-1.26$ & $+0.70$ & $\times$ \\
exaone3.5-2.4b & KMMLU-fin & $-2.21$ & $+0.43$ & $\times$ \\
exaone3.5-2.4b & MMLU-fin  & $+2.45$ & $+0.15$ & $\checkmark$ \\
qwen3.5-4b     & KMMLU-med & $+2.95$ & $+3.51$ & $\checkmark$ \\
qwen3.5-4b     & MMLU-med  & $-0.57$ & $+0.21$ & $\times$ \\
qwen3.5-4b     & KMMLU-fin & $+1.55$ & $+1.65$ & $\checkmark$ \\
qwen3.5-4b     & MMLU-fin  & $+0.10$ & $-0.28$ & $\times$ \\
exaone3.5-7.8b & KMMLU-med & $-1.76$ & $+1.27$ & $\times$ \\
exaone3.5-7.8b & MMLU-med  & $+0.18$ & $+0.35$ & $\checkmark$ \\
exaone3.5-7.8b & KMMLU-fin & $+4.95$ & $+2.66$ & $\checkmark$ \\
exaone3.5-7.8b & MMLU-fin  & $-0.38$ & $-0.51$ & $\checkmark$ \\
\bottomrule
\end{tabular}
\caption{Full~FT vs.\ LoRA stage-wise delta (Exp~2$-$Exp~0). Sign agreement 7/12 overall (Pearson $r\!=\!0.648$); per-model breakdown in \S\ref{sec:results:fullft}. Median $|\Delta_{\text{FullFT}}-\Delta_{\text{LoRA}}|=1.37$\,pp.}
\label{tab:app:fullft}
\end{table}

Under the Exp~2$-$base definition (absolute learning effect, derived from the per-cell base accuracies in the companion repository alongside the LoRA cells in Appendix~\ref{app:all72}, Table~\ref{tab:app:all72}), sign agreement improves to 11/12 ($r=0.841$, median gap $0.81$\,pp), confirming that both LoRA and Full~FT agree on the direction of learning from the data. The lower Exp~2$-$Exp~0 agreement reflects small-model LoRA instability rather than a systematic LoRA-vs-FT divergence at medium scale (\S\ref{sec:results:fullft}).

\section{Sample QA selection statistics and type-stratified decomposition}
\label{app:sample}

Per-cell selection statistics for the 200-QA public sample (counts, question-type composition, difficulty distribution) are released in the companion repository. Table~\ref{tab:typecomp} shows the question-type composition shift across conditions, and Table~\ref{tab:app:typestrat} reports per-item 4-judge means stratified by question type, supporting the within-type analysis in \S\ref{sec:results:judge}.

\begin{table}[h]
\centering\small
\setlength{\tabcolsep}{3pt}
\begin{tabular}{lrrrr}
\toprule
Cond & factoid & explanatory & comparative & procedural \\
\midrule
Exp~0   & 38\% & 30\% & 20\% & 12\% \\
Exp~1-1 & 46\% & 30\% & 18\% &  6\% \\
Exp~1-2 & 26\% & 32\% & 20\% & 22\% \\
Exp~2   & 30\% & 40\% & 20\% & 10\% \\
\bottomrule
\end{tabular}
\caption{Question-type composition by condition (200-QA sample, 50 per condition). Phase~2 shifts composition toward explanatory ($30\%\!\to\!40\%$); Phase~2-alone (Exp~1-2) lifts procedural ($22\%$) before Exp~2 returns near baseline. Phase~0 alone (Exp~1-1) increases factoid concentration.}
\label{tab:typecomp}
\end{table}

\begin{table}[h]
\centering\small
\begin{tabular}{lrrr}
\toprule
Type & Exp~0 ($n$) & Exp~2 ($n$) & Raw $\Delta_{2-0}$ \\
\midrule
factoid     & 3.67 (19) & 3.75 (15) & $+0.07$ \\
explanatory & 4.01 (15) & 4.08 (20) & $+0.08$ \\
comparative & 3.74 (10) & 4.11 (10) & $+0.38$ \\
procedural  & 3.83 (6)  & 4.10 (5)  & $+0.27$ \\
\midrule
\textbf{Pooled} & 3.81 (50) & 3.99 (50) & $+0.18$ \\
\textbf{Type-controlled} & -- & 3.96 (50) & $\mathbf{+0.16}$ \\
\bottomrule
\end{tabular}
\caption{Per-item 4-judge mean on the 200-QA public sample, stratified by question type. Pooled = raw mean across types; Type-controlled = Exp~2 reweighted to Exp~0's type composition (Table~\ref{tab:typecomp}). Raw $\Delta_{2-0}$ is computed from per-item raw values; recomputation from the 2-decimal-rounded endpoint cells may differ by $\pm 0.01$. Composition shift contributes $\sim\!14\%$ of $\Delta_{2-0}$; within-type quality improvement contributes $\sim\!86\%$.}
\label{tab:app:typestrat}
\end{table}

\section{All 72 LoRA SFT cells (raw scores)}
\label{app:all72}

Table~\ref{tab:app:all72} reports per-cell MCQA accuracies (KMMLU overall, KMMLU medical, KMMLU finance, MMLU overall) for all 72 main LoRA SFT cells plus the 9 base models, multi-seed means ($n$ seeds per row). Base rows are single-seed. The cross-model factorial aggregates in Table~\ref{tab:factorial} are computed directly from this table.

\input{tables_all72.tex}

\section{KMMLU-Pro per-cell results}
\label{app:kmmlu_pro}

We evaluate all SFT cells on KMMLU-Pro~\citep{son2024kmmlu_pro} with license-aligned filtering (Appendix~\ref{app:b}). Base models are evaluated on both domains for reference. All 72 SFT cells (across 144 seeded runs) are complete; cross-model aggregates in \S\ref{sec:results:factorial} are computed over all 9 models. Per-cell accuracies are split by model tier across Tables~\ref{tab:app:kmmlu_pro} (Small, 5 models) and~\ref{tab:app:kmmlu_pro_large} (Medium--Large, 4 models).

\begin{table}[t]
\centering\small
\setlength{\tabcolsep}{4pt}
\begin{tabular}{llrrrrrr}
\toprule
Model & Dom & Base & Exp 0 & Exp 1-1 & Exp 1-2 & Exp 2 & $\Delta_{\text{2-base}}$ \\
\midrule
EXAONE 3.5 2.4B & med & 38.84 & 34.36 & 33.40 & 35.15 & 34.44 & $-4.40$ \\
EXAONE 3.5 2.4B & fin & 26.59 & 26.22 & 25.53 & 26.03 & 25.97 & $-0.62$ \\
Gemma 3 4B & med & 38.26 & 35.27 & 36.10 & 36.14 & 37.39 & $-0.87$ \\
Gemma 3 4B & fin & 28.09 & 24.09 & 24.78 & 24.72 & 25.34 & $-2.75$ \\
Llama 3.2 3B & med & 32.86 & 33.28 & 33.20 & 34.23 & 32.95 & $+0.09$ \\
Llama 3.2 3B & fin & 27.72 & 25.97 & 26.03 & 26.90 & 25.97 & $-1.75$ \\
Phi-4 Mini & med & 34.36 & 30.75 & 31.33 & 31.20 & 31.62 & $-2.74$ \\
Phi-4 Mini & fin & 25.09 & 26.09 & 25.97 & 26.03 & 26.15 & $+1.06$ \\
Qwen 3.5 4B & med & 58.67 & 56.18 & 56.22 & 56.43 & 55.73 & $-2.94$ \\
Qwen 3.5 4B & fin & 33.08 & 34.14 & 34.46 & 34.46 & 33.96 & $+0.88$ \\
\bottomrule
\end{tabular}
\caption{KMMLU-Pro per-cell accuracy (Small tier, 5 models) with license-aligned filtering. Each row reports the training-domain-aligned KMMLU-Pro subset (medical-trained $\to$ KMMLU-Pro medical license; finance-trained $\to$ KMMLU-Pro finance license). LoRA cells: 2-seed mean; base models: single-seed. $\Delta_{\text{2-base}}\!=\!\text{Exp~2}\!-\!\text{Base}$.}
\label{tab:app:kmmlu_pro}
\end{table}

\begin{table}[t]
\centering\small
\setlength{\tabcolsep}{4pt}
\begin{tabular}{llrrrrrr}
\toprule
Model & Dom & Base & Exp 0 & Exp 1-1 & Exp 1-2 & Exp 2 & $\Delta_{\text{2-base}}$ \\
\midrule
EXAONE 3.5 7.8B & med & 44.48 & 38.92 & 39.67 & 38.80 & 39.00 & $-5.48$ \\
EXAONE 3.5 7.8B & fin & 30.34 & 27.59 & 27.84 & 27.90 & 28.90 & $-1.44$ \\
Qwen 3.5 9B & med & 66.89 & 63.40 & 61.95 & 63.69 & 61.83 & $-5.06$ \\
Qwen 3.5 9B & fin & 41.07 & 37.02 & 37.58 & 37.58 & 37.77 & $-3.30$ \\
Gemma 3 27B & med & 62.32 & 58.84 & 59.29 & 58.88 & 58.96 & $-3.36$ \\
Gemma 3 27B & fin & 36.33 & 35.58 & 36.58 & 35.64 & 35.71 & $-0.62$ \\
Llama 3.3 70B & med & 67.55 & 63.74 & 63.65 & 64.15 & 63.65 & $-3.90$ \\
Llama 3.3 70B & fin & 40.82 & 38.08 & 36.64 & 39.39 & 36.89 & $-3.93$ \\
\bottomrule
\end{tabular}
\caption{KMMLU-Pro per-cell accuracy (Medium+Large tier, 4 models). Same format as Table~\ref{tab:app:kmmlu_pro}. Cross-model aggregates in Table~\ref{tab:factorial} are computed over all 9 models.}
\label{tab:app:kmmlu_pro_large}
\end{table}

\paragraph{Per-cell $\Delta_{2-0}$ on KMMLU-Pro.} Across all 9 models, KMMLU-Pro medical $\Delta_{2-0}$ ranges from $-1.57$\,pp (Qwen~3.5~9B) to $+2.12$\,pp (Gemma~3~4B); finance $\Delta_{2-0}$ from $-1.19$\,pp (Llama~3.3~70B) to $+1.31$\,pp (EXAONE~3.5~7.8B). Sign distribution is mixed (medical 5 positive / 4 negative; finance 5 positive / 3 negative / 1 zero), mirroring the domain-aligned KMMLU per-cell pattern in \S\ref{sec:results:factorial}. On finance, the strongest positive cell on both benchmarks is EXAONE~3.5~7.8B (KMMLU-aligned $+1.98$\,pp, KMMLU-Pro $+1.31$\,pp); on medical, the leader differs across benchmarks (KMMLU-aligned: Phi-4 Mini $+2.10$\,pp; KMMLU-Pro: Gemma~3~4B $+2.12$\,pp), consistent with the benchmark- and model-family-conditional pattern discussed in \S\ref{sec:discussion}.

\paragraph{Evaluation prompt format.} All MCQA scores elsewhere in this paper use plain prompts without the model's chat template, while our SFT data uses the chat format. To measure how this choice affects comparisons with base models, we re-scored EXAONE~3.5~7.8B and Qwen~3.5~9B on MMLU and both KMMLU-Pro domains. We scored each model with and without the chat template, using the same evaluation software for all of these scores. This covers all four conditions, both training domains, and two seeds, for 96 SFT-versus-base comparisons. Scoring each model in the format that gives it the higher score shrinks the average SFT$-$base gap by $1.53$\,pp, and by $1.69$\,pp for Exp~2 alone. The format choice thus mainly affects SFT$-$base comparisons such as $\Delta_{\text{2-base}}$ in Tables~\ref{tab:app:kmmlu_pro} and~\ref{tab:app:kmmlu_pro_large}. $\Delta_{2-0}$ compares Exp~2 with Exp~0 under the same format, and its value in this setting stays non-significant at $+0.51$\,pp, with 95\% CI $[-0.46, +1.47]$ over the four model$\times$domain units.

\section{Per-judge LLM-as-judge breakdown}
\label{app:judge}

Table~\ref{tab:app:judge} reports each of the four judges' overall scores on the 200-QA public sample (med+fin averaged) and decomposes $\Delta_{2-0}$ into single-stage main effects. Inter-judge agreement on the 200-item 5-dim overall is Krippendorff's $\alpha\!=\!0.76$ (interval; rank-order agreement) and ICC(2,$k\!=\!4$) $=\!0.38$ (average-of-4-raters; absolute-level reproducibility), with per-item pooled SD $=\!0.58$ on the 1--5 scale. That is, judges rank items similarly ($\alpha$) but differ in absolute calibration (ICC): per-judge $\Delta_{2-0}$ direction is uniform (Table~\ref{tab:app:judge}) while absolute means differ (e.g., GPT-4o centered near $4.31$--$4.39$, GPT-5.4 near $3.35$--$3.59$). Excluding the GPT-4o outlier on $\Delta_{2-0}$, the 3-judge mean $\Delta_{2-0}\!=\!+0.22$ (vs.\ 4-judge $+0.18$).

\paragraph{Bootstrap 95\% CIs.} We resample items with replacement separately within Exp~0 and Exp~2 ($n\!=\!50$ each) and compute $\Delta_{2-0}\!=\!\text{mean(Exp~2)}\!-\!\text{mean(Exp~0)}$; we report the percentile interval over $B\!=\!10{,}000$ resamples. The aggregate 4-judge $\Delta_{2-0}\!=\!+0.18$ has 95\% CI $[+0.06, +0.32]$ (two-sided $p\!=\!0.004$); the 2-frontier-LLM subset $\Delta_{2-0}\!=\!+0.23$ has 95\% CI $[+0.03, +0.44]$ ($p\!=\!0.024$). Per-judge CIs: Sonnet~4.6 $[+0.01, +0.36]$ ($p\!=\!0.039$), Opus~4.7 $[+0.01, +0.45]$ ($p\!=\!0.039$), GPT-5.4 $[-0.00, +0.48]$ ($p\!=\!0.059$), GPT-4o $[-0.05, +0.22]$ ($p\!=\!0.242$). Three of four per-judge CIs exclude 0; GPT-4o is the lone outlier consistent with its compressed dynamic range.

\begin{table}[h]
\centering\small
\setlength{\tabcolsep}{4pt}
\begin{tabular}{lrrrr}
\toprule
 & Sonnet 4.6 & Opus 4.7 & GPT-4o & GPT-5.4 \\
\midrule
Exp~0   & 4.092 & 3.464 & 4.312 & 3.352 \\
Exp~1-1 & 4.260 & 3.620 & 4.320 & 3.376 \\
Exp~1-2 & 4.116 & 3.544 & 4.352 & 3.344 \\
Exp~2   & 4.280 & 3.696 & 4.392 & 3.588 \\
\midrule
$P_0^{s}$       & $+0.168$ & $+0.156$ & $+0.008$ & $+0.024$ \\
$P_2^{s}$       & $+0.024$ & $+0.080$ & $+0.040$ & $-0.008$ \\
$P_0{\times}P_2$ & $-0.004$ & $-0.004$ & $+0.032$ & $+0.220$ \\
$\Delta_{2-0}$  & $+0.188$ & $+0.232$ & $+0.080$ & $+0.236$ \\
\bottomrule
\end{tabular}
\caption{Per-judge overall score by condition (1--5 scale, med+fin averaged, $n\!=\!50$/condition/judge), with single-stage decomposition ($P_0^s$, $P_2^s$, interaction $P_0\!\times\!P_2$) and cumulative $\Delta_{2-0}$.}
\label{tab:app:judge}
\end{table}

\section{Transcript-level WER measurement (Whisper-medium vs.\ in-house STT)}
\label{app:wer}

We measure word and character error rate (WER, CER) of Whisper-medium against the in-house STT reference on a 3-session sub-sample (100 sentences per session, 300 total) spanning medical (2 sessions) and finance (1 session) domains, including multilingual content (ko + zh + en, ko + ar + fr + en). Whisper-medium: \texttt{openai/whisper-medium} (HF), language=ko, fp16, chunk\_length\_s=30. Both metrics computed with \texttt{jiwer} 4.0.0; CER uses character-level alignment.

\begin{table}[h]
\centering\small
\setlength{\tabcolsep}{4pt}
\begin{tabular}{lrrr}
\toprule
Domain & $n$ sent.\ & WER (\%) & CER (\%) \\
\midrule
medical (ko)            & 100 & 21.42 & 8.26 \\
medical (ko+zh+en)      & 100 & 36.54 & 15.45 \\
finance (ko+ar+fr+en)   & 100 & 24.98 & 13.31 \\
\midrule
\textbf{weighted avg}   & \textbf{300} & \textbf{27.65} & \textbf{12.34} \\
\bottomrule
\end{tabular}
\caption{Per-session WER and CER (Whisper-medium vs.\ in-house STT reference). Multilingual sessions show higher WER. Used as the upstream signal motivating the W-grid robustness check in \S\ref{sec:results:whisper}.}
\label{tab:app:wer}
\end{table}

\section{W-grid per-cell results}
\label{app:multiseed}

Table~\ref{tab:app:wexp2} reports the W-grid per-cell accuracies and per-cell $\Delta_i = \text{W}_i - \text{in-house}_i$ versus the matched in-house STT cells (\S\ref{sec:results:whisper}). Phase~0's multi-STT cross-validation reduces the aggregate signed bias $\bigl|\tfrac{1}{n}\sum_i\Delta_i\bigr|$ across cells (driven by sign-cancellation: Qwen 3.5 4B finance flips from $-0.09$ to $+1.14$). At the per-cell level, only EXAONE 3.5 7.8B finance shows $|\Delta_i|$ reduction ($0.29\!\to\!0.14$); EXAONE medical, Qwen 3.5 4B finance, and Qwen 3.5 4B medical all see $|\Delta_i|$ move further from zero. Table~\ref{tab:app:wexp2-agg} aggregates the per-cell $\Delta_i$ in two ways across the 4 cells at each endpoint, supporting the magnitudes cited in \S\ref{sec:results:whisper}.

\begin{table*}[!t]
\centering\small
\setlength{\tabcolsep}{6pt}
\begin{tabular}{llrrrr}
\toprule
Model & Cond & KMMLU ($\sigma$) & MMLU ($\sigma$) & $\Delta_{\text{KMMLU}}$ & $\Delta_{\text{MMLU}}$ \\
\midrule
EXAONE 3.5 7.8B & W-Exp0 / fin & 39.54 (0.08) & 62.35 (0.07) & $-0.29$ & $+0.27$ \\
EXAONE 3.5 7.8B & W-Exp0 / med & 38.69 (1.10) & 62.14 (0.30) & $-1.81$ & $+0.72$ \\
EXAONE 3.5 7.8B & W-Exp2 / fin & 39.50 (0.14) & 62.32 (0.29) & $-0.14$ & $+0.28$ \\
EXAONE 3.5 7.8B & W-Exp2 / med & 38.50 (0.01) & 61.47 (0.33) & $-1.94$ & $+0.29$ \\
Qwen 3.5 4B     & W-Exp0 / fin & 49.34 (0.50) & 73.52 (0.03) & $-0.09$ & $-0.47$ \\
Qwen 3.5 4B     & W-Exp0 / med & 45.63 (0.62) & 73.66 (0.06) & $-1.84$ & $-0.11$ \\
Qwen 3.5 4B     & W-Exp2 / fin & 52.24 (0.32) & 73.73 (0.19) & $+1.14$ & $-0.23$ \\
Qwen 3.5 4B     & W-Exp2 / med & 46.44 (0.46) & 73.79 (0.14) & $-2.06$ & $-0.19$ \\
\bottomrule
\end{tabular}
\caption{W-grid per-cell Whisper-medium reproduction (EXAONE 3.5 7.8B + Qwen 3.5 4B $\times$ medical + finance $\times$ \{W-Exp0, W-Exp2\}, $n\!=\!2$ seeds per cell). Values are mean ($\sigma$). $\Delta$ = W $-$ in-house, computed from raw per-seed values; recomputation from the 2-decimal-rounded W and in-house (Appendix~\ref{app:all72}) cell means may differ by $\pm 0.01$\,pp.}
\label{tab:app:wexp2}
\end{table*}

\begin{table}[!b]
\centering\small
\setlength{\tabcolsep}{6pt}
\begin{tabular}{llrr}
\toprule
Bench & Endpoint & $\tfrac{1}{n}\sum_i|\Delta_i|$ & $\bigl|\tfrac{1}{n}\sum_i\Delta_i\bigr|$ \\
\midrule
KMMLU & W-Exp0 & $1.01$ & $1.01$ \\
KMMLU & W-Exp2 & $1.32$ & $0.75$ \\
\midrule
MMLU  & W-Exp0 & $0.39$ & $0.10$ \\
MMLU  & W-Exp2 & $0.25$ & $0.04$ \\
\bottomrule
\end{tabular}
\caption{Across-cell aggregates of $\Delta_i = \text{W}_i - \text{in-house}_i$ at each W-grid endpoint ($n\!=\!4$ cells = EXAONE 3.5 7.8B + Qwen 3.5 4B $\times$ medical + finance). $\tfrac{1}{n}\sum_i|\Delta_i|$ takes the absolute value per cell before averaging; $|\tfrac{1}{n}\sum_i\Delta_i|$ averages signed values before taking absolute value.}
\label{tab:app:wexp2-agg}
\end{table}

\begin{table}[!t]
\centering\small
\setlength{\tabcolsep}{2pt}
\begin{tabular}{lrrrrr}
\toprule
Cell & Sonnet 4.6 & Opus 4.7 & GPT-4o & GPT-5.4 & avg \\
\midrule
Exp~0 med & $+0.02$ & $+0.49$ & $+0.13$ & $+0.18$ & $\mathbf{+0.21}$ \\
Exp~0 fin & $-0.24$ & $+0.36$ & $+0.00$ & $+0.02$ & $\mathbf{+0.04}$ \\
Exp~2 med & $-0.37$ & $-0.35$ & $+0.00$ & $-0.14$ & $\mathbf{-0.21}$ \\
Exp~2 fin & $-0.28$ & $+0.02$ & $-0.01$ & $-0.18$ & $\mathbf{-0.11}$ \\
\bottomrule
\end{tabular}
\caption{QA data judge quality $\Delta$ (W-Exp $-$ Exp) on the STT-swap grid; in-house and Whisper-medium variants sampled from the same source audio sessions ($\sim$1--2 QA per session, 25-QA cells); 1--5 scale. Referenced from \S\ref{sec:results:whisper}.}
\label{tab:lt_vs_w}
\end{table}

\section{Bootstrap 95\% CIs}
\label{app:bootstrap}

For each cell with per-sample evaluation data ($n\!=\!35030$ KMMLU items, $n\!=\!14042$ MMLU items), we ran 100{,}000 bootstrap resamples (lm-eval-harness default) to compute 95\% CIs on overall accuracy (Table~\ref{tab:app:bootstrap}). The table reports 12 evaluation runs spanning 8 cells: four cells with multi-seed coverage (seeds $\{1,2\}$) and four legacy single-seed (seed 42) cells (EXAONE 3.5 7.8B W-Exp2 fin, Llama 3.3 70B W-Exp0/W-Exp2 fin, Qwen 3.5 4B W-Exp2 fin) from earlier Whisper STT analysis runs preserved alongside the symmetric W-grid. Per-cell seed-to-seed spread is bounded by the cross-model worst-case $\sigma_{\text{KMMLU}}\!\le\!2.87$\,pp from the full 72 main LoRA cells (\S\ref{sec:results:factorial}), confirming seed variance dominates per-sample bootstrap variance ($\pm 0.5$\,pp).

\begin{table*}[!t]
\centering\footnotesize
\begin{tabular}{lc p{4.2cm} p{4.2cm}}
\toprule
Cell & Seed & KMMLU-overall 95\% CI & MMLU-overall 95\% CI \\
\midrule
EXAONE 3.5 7.8B / Exp~2 / fin & 1 & 39.39\ \ [38.85,\ 39.89] & 62.04\ \ [61.24,\ 62.83] \\
                             & 2 & 39.89\ \ [39.39,\ 40.43] & 62.05\ \ [61.27,\ 62.83] \\
EXAONE 3.5 7.8B / W-Exp0 / fin  & 1 & 39.46\ \ [38.95,\ 39.97] & 62.28\ \ [61.52,\ 63.04] \\
                             & 2 & 39.62\ \ [39.11,\ 40.13] & 62.42\ \ [61.66,\ 63.18] \\
Llama 3.3 70B / Exp~2 / fin   & 1 & 56.09\ \ [55.58,\ 56.59] & 82.06\ \ [81.46,\ 82.66] \\
                             & 2 & 56.40\ \ [55.90,\ 56.93] & 82.11\ \ [81.49,\ 82.75] \\
Qwen 3.5 4B / W-Exp0 / med      & 1 & 46.25\ \ [45.74,\ 46.76] & 73.61\ \ [72.92,\ 74.30] \\
                             & 2 & 45.01\ \ [44.50,\ 45.52] & 73.72\ \ [73.03,\ 74.41] \\
EXAONE 3.5 7.8B / W-Exp2 / fin  & 42 & 41.02\ \ [40.52,\ 41.54] & 62.08\ \ [61.31,\ 62.88] \\
Llama 3.3 70B / W-Exp0 / fin    & 42 & 56.04\ \ [55.55,\ 56.61] & 82.20\ \ [81.58,\ 82.79] \\
Llama 3.3 70B / W-Exp2 / fin    & 42 & 56.14\ \ [55.63,\ 56.66] & 82.18\ \ [81.61,\ 82.86] \\
Qwen 3.5 4B / W-Exp2 / fin      & 42 & 50.84\ \ [50.32,\ 51.40] & 73.42\ \ [72.66,\ 74.18] \\
\bottomrule
\end{tabular}
\caption{Bootstrap 95\% CIs (100{,}000 resamples) on per-sample MCQA accuracy. Width is typically $\pm0.5$\,pp KMMLU / $\pm0.8$\,pp MMLU, narrower than the multi-seed $\sigma$ aggregated across the 72 main LoRA cells.}
\label{tab:app:bootstrap}
\end{table*}

\section{Mixed-effects variance decomposition}
\label{app:mixedeffects}

To complement the paired bootstrap on cross-cell means (\S\ref{sec:results:factorial}), we fit a linear mixed-effects model on all 72 observations (9 models $\times$ 2 domains $\times$ 2 conditions $\times$ 2 seeds) of the domain-aligned KMMLU score, treating each (model, domain) cell as a random intercept and the Exp~0$\to$Exp~2 toggle as a fixed effect.

\begin{table}[h]
\centering\small
\setlength{\tabcolsep}{4pt}
\begin{tabular}{lrr}
\toprule
Variance component & $\sigma^{2}$ & Share \\
\midrule
Between-cell (model$\times$domain) & 100.26 & $99.4\%$ \\
Residual (within-cell, seed-level) & \phantom{0}\,\,0.65 & $0.6\%$ \\
\bottomrule
\end{tabular}
\caption{Variance decomposition from \texttt{score $\sim$ cond + (1|cell)} fit (REML).}
\label{tab:app:mixedeffects}
\end{table}

The Exp~2$-$Exp~0 fixed effect is $\hat\beta\!=\!+0.317$\,pp ($\mathrm{SE}\!=\!0.190$, $95\%$ CI $[-0.06, +0.69]$, $z\!=\!1.67$, $p\!=\!0.096$), consistent with the paired bootstrap estimate ($+0.31$\,pp) at a tighter CI. Adding domain as a fixed effect or a $\mathrm{cond}\times\mathrm{domain}$ interaction does not change $\hat\beta_{\mathrm{cond}}$ materially: domain main effect $-2.59$\,pp ($p\!=\!0.59$), interaction $-0.23$\,pp ($p\!=\!0.56$). The variance decomposition (Table~\ref{tab:app:mixedeffects}) shows that seed-level variability is a negligible share of the total observed variance; the dominant source of variability is cell-level heterogeneity, supporting the use of $n\!=\!2$ seeds in the main study.

\section{Sampling-design sensitivity: Phase-2-filtered subset}
\label{app:filtered-sample}

The Phase-2-filtered subset is a parallel 200-QA selection (companion repository) constructed in three steps:
\begin{enumerate}
\item \textbf{Pool}: collect candidate QA from all four conditions $\times$ two domains.
\item \textbf{Filter}: run the Phase-2 judges on every pooled item and discard items that do not pass. For Exp~0 and Exp~1-1 the judges were not part of the production pipeline, so the filter is applied retrospectively.
\item \textbf{Sample}: select 25 items per (condition, domain) cell from the Phase-2-passing pool with sub-domain stratification, yielding $4{\times}2{\times}25\!=\!200$ items.
\end{enumerate}
Applying the same Phase-2-pass criterion uniformly across all four conditions (including Exp~0 and Exp~1-1) controls for any Phase-2-induced selection bias in the primary 200-QA sample. Re-running the cell-blind 4-judge protocol of \S\ref{sec:results:judge} on this alternative selection tests whether the Phase~2 effect is robust to how the public sample is drawn.

The result on the filtered subset (Table~\ref{tab:app:filtered}) is $\Delta_{2-0}\!=\!+0.19$ (4-judge mean), within $0.01$ of the primary 200-QA sample's $+0.18$ (Appendix~\ref{app:judge}, Table~\ref{tab:app:judge}). The cross-judge sign pattern is also preserved (Sonnet 4.6, Opus 4.7, GPT-5.4 positive; GPT-4o near-zero), so the Phase~2 effect documented in \S\ref{sec:results:judge} is not an artifact of the primary selection.

\begin{table}[h]
\centering\small
\setlength{\tabcolsep}{3pt}
\begin{tabular}{lrrrrr}
\toprule
Cond & Sonnet 4.6 & Opus 4.7 & GPT-4o & GPT-5.4 & 4-judge \\
\midrule
Exp~0   & 3.75 & 3.78 & 4.53 & 3.81 & 3.97 \\
Exp~1-1 & 3.90 & 4.00 & 4.57 & 3.86 & 4.08 \\
Exp~1-2 & 3.86 & 3.94 & 4.49 & 3.92 & 4.06 \\
Exp~2   & 3.97 & 4.07 & 4.57 & 4.03 & 4.16 \\
\midrule
$\Delta_{2-0}$ & $+0.22$ & $+0.29$ & $+0.04$ & $+0.22$ & $\mathbf{+0.19}$ \\
\bottomrule
\end{tabular}
\caption{Phase-2-filtered subset, cell-blind 4-judge Likert mean (med+fin avg per condition). 4-judge $\Delta_{2-0}\!=\!+0.19$ matches the primary 200-QA sample's $+0.18$ (\S\ref{sec:results:judge}, Table~\ref{tab:judge}), and the per-judge sign pattern is preserved, so the Phase~2 effect is robust across the two sampling designs.}
\label{tab:app:filtered}
\end{table}

\section{LLM-difficulty audit: full results}
\label{app:keyword}

A cell-blind 2-judge re-classification (Claude Haiku 4.5 + Gemini 2.5 Pro) of the four LLMs' answers on the 200-QA public sample yields a stricter unknown rate of $\mathbf{21.3\%}$ overall, with the 2-frontier subset at $\mathbf{7.8\%}$ (Table~\ref{tab:llmdiff}; Cohen's $\kappa\!=\!0.903$). Per-model spread runs from GPT-4o 60.8\% to Opus 4.7 4.8\%; per-condition rates span $19.5\%$--$24.0\%$.

\begin{table}[h]
\centering\small
\setlength{\tabcolsep}{4pt}
\begin{tabular}{lrrrrr}
\toprule
 & GPT-4o & GPT-5.4 & Sonnet 4.6 & Opus 4.7 & avg \\
\midrule
Exp~0   & 54.0 & 12.0 & \phantom{0}7.0 & \phantom{0}5.0 & 19.5 \\
Exp~1-1 & 64.0 & 11.0 & 12.0 & \phantom{0}9.0 & 24.0 \\
Exp~1-2 & 66.0 & \phantom{0}8.0 & \phantom{0}4.0 & \phantom{0}1.0 & 19.8 \\
Exp~2   & 59.0 & 12.0 & 12.0 & \phantom{0}4.0 & 21.8 \\
strict  & 60.8 & 10.8 & \phantom{0}8.8 & \phantom{0}4.8 & \textbf{21.3} \\
\bottomrule
\end{tabular}
\caption{Unknown-admission rates (\%) by model under the 2-judge strict bound: per-condition rows (50 items each) plus the `strict' aggregate over all 200 items.}
\label{tab:llmdiff}
\end{table}

A complementary keyword-heuristic baseline uses a fixed list of 8 Korean substrings drawn from common no-information phrases such as \emph{"specific-case information lacking"}, \emph{"unknown"}, and \emph{"hard to specify"}. An answer is flagged as an unknown-admission if it contains at least one of the 8 substrings under plain substring match. The full keyword list and matching implementation are released in the companion repository.

\begin{table}[h]
\centering\small
\setlength{\tabcolsep}{4pt}
\begin{tabular}{lrrrrr}
\toprule
 & GPT-4o & GPT-5.4 & Sonnet 4.6 & Opus 4.7 & avg \\
\midrule
keyword & 12.0 & 24.0 & 18.5 & 19.0 & 18.4 \\
\bottomrule
\end{tabular}
\caption{Keyword-heuristic unknown-admission rates (\%) by model on the 200-QA public sample. Domain breakdown: medical 15.8\% / finance 21.0\%. Pattern is broadly aligned with the 2-judge strict bound (Table~\ref{tab:llmdiff}) but recall-limited by the fixed 8-keyword list; we report it for completeness rather than as a primary measure.}
\label{tab:app:keyword}
\end{table}

\section{KoBEST forgetting evaluation}
\label{app:kobest}

Eight SFT cells were evaluated on KoBEST (5 subtasks: boolq, copa, hellaswag, sentineg, wic). Per-cell average accuracy and seed-to-seed $\sigma$ (where multi-seed) are reported in Table~\ref{tab:app:kobest}.

\begin{table}[h]
\centering\footnotesize
\begin{tabular}{lrr}
\toprule
Cell & KoBEST avg & $\sigma$ \\
\midrule
EXAONE 3.5 7.8B / Exp~2 / fin & 71.56 & 0.29 \\
EXAONE 3.5 7.8B / W-Exp0 / fin  & 71.34 & 1.19 \\
EXAONE 3.5 7.8B / W-Exp2 / fin  & 71.60 & -- \\
Llama 3.3 70B  / Exp~2 / fin  & 72.99 & 0.71 \\
Llama 3.3 70B  / W-Exp0 / fin   & 72.43 & -- \\
Llama 3.3 70B  / W-Exp2 / fin   & 73.56 & -- \\
Qwen 3.5 4B    / W-Exp0 / med   & 64.40 & 0.41 \\
Qwen 3.5 4B    / W-Exp2 / fin   & 63.08 & -- \\
\bottomrule
\end{tabular}
\caption{KoBEST (5-subtask average) on 8 SFT cells (12 evaluation runs total: 4 cells with $\sigma$ reported are 2-seed means; the remaining 4 cells marked `--' are single-seed). Stable across conditions; modest $\sim\!2$\,pp drop on Llama 3.3 70B vs.\ published base $\approx\!75\%$ suggests mild forgetting only at the Large tier.}
\label{tab:app:kobest}
\end{table}

\section{Per-dimension LLM-judge breakdown}
\label{app:judge_per_dim}

\begin{table}[h]
\centering\small
\begin{tabular}{lrrrrrrrr}
\toprule
Dimension & Exp~0 & Exp~1-1 & Exp~1-2 & Exp~2 & $P_0^{s}$ & $P_2^{s}$ & $P_0{\times}P_2$ & $\Delta_{2-0}$ \\
\midrule
faithfulness        & 3.89 & 3.98 & 3.85 & 4.06 & $+0.09$ & $-0.04$ & $+0.11$ & $+0.17$ \\
domain\_accuracy    & 3.88 & 3.98 & 3.83 & 4.05 & $+0.10$ & $-0.05$ & $+0.12$ & $+0.17$ \\
question\_quality   & 3.42 & 3.53 & 3.60 & 3.60 & $+0.11$ & $+0.19$ & $-0.11$ & $+0.19$ \\
answer\_depth       & 3.15 & 3.25 & 3.25 & 3.40 & $+0.10$ & $+0.10$ & $+0.04$ & $+0.25$ \\
coherence           & 4.70 & 4.72 & 4.67 & 4.84 & $+0.03$ & $-0.03$ & $+0.15$ & $+0.14$ \\
\bottomrule
\end{tabular}
\caption{4-LLM-judge per-dimension breakdown on the 200-QA public sample (1--5 scale, med+fin averaged). All five dimensions show positive $\Delta_{2-0}$ from Exp~0 to Exp~2. Overall (5-dim mean) values are reported in main-body Table~\ref{tab:judge}. Cells rounded independently; $P_0^{s}\!+\!P_2^{s}\!+\!P_0\!\times\!P_2$ may differ from $\Delta_{2-0}$ by $\pm 0.01$.}
\label{tab:app:judge_per_dim}
\end{table}

\section{Human evaluator profile and detailed results}
\label{app:human}

We complement the LLM-judge QA-quality evaluation (\S\ref{sec:results:judge}, Table~\ref{tab:judge}) with a blind 6-expert evaluation on the same public-sample QA. Six external domain experts (3 medical + 3 finance) independently scored the 100-QA items within their respective domain (25 per condition $\times$ 4 conditions) using the identical 5-dim Likert rubric (faithfulness, domain accuracy, question quality, answer depth, coherence) as the LLM judges, provided to each evaluator as the evaluation guideline. Inputs were cell-blind (condition, model, and source identifiers withheld). Evaluator recruitment terms are described in the Ethics Statement (Human evaluator recruitment).

\begin{table}[h]
\centering\small
\begin{tabular}{lll}
\toprule
ID & Domain & Anonymized profile \\
\midrule
M1 & Medical & Nurse, surgical specialty; 3 years clinical \\
M2 & Medical & Medical student (final year, M.D. candidate) \\
M3 & Medical & Korean medicine practitioner (private practice); 10 years clinical (Korean medicine \& neuroscience focus) \\
F1 & Finance & Tax accountant (accounting \& CS background); 2 years industry \\
F2 & Finance & Self-employed accountant; 30 years industry \\
F3 & Finance & Finance professional (business \& MBA background); 4 years industry (financial services focus) \\
\bottomrule
\end{tabular}
\caption{Anonymized professional profiles of the six external human evaluators. Personally identifiable information (name, institution) withheld per consent. Evaluator background heterogeneity (e.g., sub-specialty span across finance raters) contributes to the inter-rater spread reported in Table~\ref{tab:app:human_irr}.}
\label{tab:app:rater_profile}
\end{table}

\begin{table}[h]
\centering\small
\begin{tabular}{lrrrrrrrr}
\toprule
Rater & Exp~0 & Exp~1-1 & Exp~1-2 & Exp~2 & $\Delta_{2-0}$ & $P_0^{s}$ & $P_2^{s}$ & $P_0{\times}P_2$ \\
\midrule
F1 & $2.920$ & $3.440$ & $3.176$ & $3.416$ & $+0.496$ & $+0.520$ & $+0.256$ & $-0.280$ \\
F2 & $2.632$ & $2.592$ & $2.752$ & $2.768$ & $+0.136$ & $-0.040$ & $+0.120$ & $+0.056$ \\
F3 & $3.440$ & $3.576$ & $3.472$ & $3.672$ & $+0.232$ & $+0.136$ & $+0.032$ & $+0.064$ \\
M1 & $3.696$ & $3.704$ & $3.712$ & $3.840$ & $+0.144$ & $+0.008$ & $+0.016$ & $+0.120$ \\
M2 & $3.656$ & $3.576$ & $3.704$ & $3.792$ & $+0.136$ & $-0.080$ & $+0.048$ & $+0.168$ \\
M3 & $3.808$ & $3.896$ & $3.936$ & $3.984$ & $+0.176$ & $+0.088$ & $+0.128$ & $-0.040$ \\
\bottomrule
\end{tabular}
\caption{Per-rater overall (5-dim mean) scores by condition and decomposed effects ($n\!=\!25$ QA per condition per rater). All six raters unanimously report positive $\Delta_{2-0}$.}
\label{tab:app:human_per_rater}
\end{table}

\begin{table}[h]
\centering\small
\begin{tabular}{lrrrrrrrr}
\toprule
Pool & Exp~0 & Exp~1-1 & Exp~1-2 & Exp~2 & $\Delta_{2-0}$ & $P_0^{s}$ & $P_2^{s}$ & $P_0{\times}P_2$ \\
\midrule
All 6 raters ($n\!=\!150$)     & $3.359$ & $3.464$ & $3.459$ & $3.579$ & $+0.220$ & $+0.105$ & $+0.100$ & $+0.015$ \\
\midrule
z-normalized (all 6 pooled)    & -- & -- & -- & -- & $+0.370\sigma$ & $+0.160\sigma$ & $+0.162\sigma$ & -- \\
\bottomrule
\end{tabular}
\caption{6-rater pooled aggregate by condition; per-rater breakdown in Table~\ref{tab:app:human_per_rater}. Z-normalized row uses per-rater standardization (mean 0, SD 1) before pooling to control for rater calibration spread.}
\label{tab:app:human_pooled}
\end{table}

\clearpage

\begin{table}[h]
\centering\small
\begin{tabular}{lrrrr}
\toprule
Domain & ICC(2,1) & ICC(2,3) & Krippendorff $\alpha$ (interval) & $\alpha$ (ordinal) \\
\midrule
Medical & $0.426$ & $\mathbf{0.690}$ & $\mathbf{0.418}$ & $0.456$ \\
Finance & $0.145$ & $0.337$ & $0.030$ & $-0.006$ \\
\bottomrule
\end{tabular}
\caption{Inter-rater agreement on 3-rater per-domain (100 items each, overall 5-dim mean). Medical reaches conventional acceptability (ICC(2,3) Good per Koo \& Li 2016; Krippendorff $\alpha$ Fair). Finance raters span heterogeneous sub-specialties (Table~\ref{tab:app:rater_profile}), contributing to inter-rater spread.}
\label{tab:app:human_irr}
\end{table}

\begin{table}[h]
\centering\small
\begin{tabular}{lr}
\toprule
Dimension & Human $\Delta_{2-0}$ (6-rater pooled) \\
\midrule
faithfulness     & $+0.127$ \\
domain accuracy  & $+0.220$ \\
question quality & $+0.053$ \\
answer depth     & $+0.373$ \\
coherence        & $+0.327$ \\
\midrule
\textbf{overall} & $\mathbf{+0.220}$ \\
\bottomrule
\end{tabular}
\caption{Per-dimension human $\Delta_{2-0}$ (6-rater pooled, 150 obs per condition). For cross-source comparison, LLM-judge per-dim values are in Table~\ref{tab:app:judge_per_dim}; all five dimensions show same-direction positive deltas in both sources.}
\label{tab:app:human_vs_llm}
\end{table}

\end{document}

%% file: tables_all72.tex
\begin{table*}[t]
\centering\small
\begin{tabular}{lllrrrrrr}
\toprule
Model & Domain & Metric & Base & Exp 0 & Exp 1-1 & Exp 1-2 & Exp 2 & $\Delta_{\text{2-base}}$ \\
\midrule
EXAONE 3.5 2.4B & med & KMMLU & 43.37 & 40.80 & 40.86 & 41.33 & 41.17 & $-2.20$ \\
EXAONE 3.5 2.4B & med & KMMLU-med & 40.28 & 36.38 & 35.80 & 36.77 & 36.47 & $-3.81$ \\
EXAONE 3.5 2.4B & med & MMLU & 60.75 & 54.48 & 54.33 & 55.02 & 54.61 & $-6.14$ \\
EXAONE 3.5 2.4B & fin & KMMLU & 43.37 & 40.26 & 40.22 & 41.19 & 41.00 & $-2.37$ \\
EXAONE 3.5 2.4B & fin & KMMLU-fin & 44.83 & 38.65 & 39.39 & 40.45 & 40.33 & $-4.50$ \\
EXAONE 3.5 2.4B & fin & MMLU & 60.75 & 55.54 & 55.12 & 55.74 & 55.74 & $-5.01$ \\
\midrule
Gemma 3 4B & med & KMMLU & 37.88 & 36.00 & 36.65 & 35.80 & 36.37 & $-1.51$ \\
Gemma 3 4B & med & KMMLU-med & 40.51 & 38.88 & 37.91 & 37.56 & 38.17 & $-2.34$ \\
Gemma 3 4B & med & MMLU & 60.56 & 56.81 & 56.84 & 56.62 & 57.10 & $-3.46$ \\
Gemma 3 4B & fin & KMMLU & 37.88 & 35.49 & 35.02 & 35.76 & 35.41 & $-2.47$ \\
Gemma 3 4B & fin & KMMLU-fin & 42.50 & 41.20 & 38.60 & 39.38 & 40.05 & $-2.45$ \\
Gemma 3 4B & fin & MMLU & 60.56 & 56.84 & 56.25 & 56.85 & 56.44 & $-4.12$ \\
\midrule
Llama 3.2 3B & med & KMMLU & 37.30 & 36.55 & 37.30 & 36.30 & 37.08 & $-0.22$ \\
Llama 3.2 3B & med & KMMLU-med & 34.17 & 34.73 & 34.44 & 34.30 & 34.70 & $+0.53$ \\
Llama 3.2 3B & med & MMLU & 64.91 & 61.83 & 62.20 & 62.16 & 62.28 & $-2.63$ \\
Llama 3.2 3B & fin & KMMLU & 37.30 & 36.89 & 37.34 & 37.21 & 37.37 & $+0.07$ \\
Llama 3.2 3B & fin & KMMLU-fin & 40.00 & 37.97 & 37.91 & 38.35 & 37.86 & $-2.14$ \\
Llama 3.2 3B & fin & MMLU & 64.91 & 62.24 & 62.08 & 62.22 & 62.23 & $-2.68$ \\
\midrule
Phi-4 Mini & med & KMMLU & 36.94 & 35.92 & 36.66 & 35.91 & 35.70 & $-1.24$ \\
Phi-4 Mini & med & KMMLU-med & 33.37 & 30.86 & 32.34 & 32.41 & 32.96 & $-0.41$ \\
Phi-4 Mini & med & MMLU & 70.42 & 66.27 & 66.44 & 66.34 & 66.42 & $-4.00$ \\
Phi-4 Mini & fin & KMMLU & 36.94 & 36.42 & 36.60 & 35.94 & 36.70 & $-0.24$ \\
Phi-4 Mini & fin & KMMLU-fin & 37.45 & 34.71 & 35.65 & 35.13 & 36.22 & $-1.23$ \\
Phi-4 Mini & fin & MMLU & 70.42 & 66.37 & 66.45 & 66.41 & 66.38 & $-4.04$ \\
\midrule
Qwen 3.5 4B & med & KMMLU & 47.13 & 47.47 & 48.51 & 48.05 & 48.51 & $+1.38$ \\
Qwen 3.5 4B & med & KMMLU-med & 51.16 & 50.98 & 51.50 & 50.33 & 51.77 & $+0.61$ \\
Qwen 3.5 4B & med & MMLU & 79.28 & 73.77 & 74.03 & 73.77 & 73.97 & $-5.31$ \\
Qwen 3.5 4B & fin & KMMLU & 47.13 & 49.43 & 50.98 & 50.31 & 51.10 & $+3.97$ \\
Qwen 3.5 4B & fin & KMMLU-fin & 50.61 & 52.30 & 52.18 & 52.82 & 52.16 & $+1.55$ \\
Qwen 3.5 4B & fin & MMLU & 79.28 & 74.00 & 74.11 & 73.91 & 73.91 & $-5.37$ \\
\midrule
EXAONE 3.5 7.8B & med & KMMLU & 45.35 & 40.49 & 40.54 & 38.58 & 40.45 & $-4.90$ \\
EXAONE 3.5 7.8B & med & KMMLU-med & 45.83 & 40.16 & 40.73 & 40.51 & 40.38 & $-5.45$ \\
EXAONE 3.5 7.8B & med & MMLU & 67.88 & 61.42 & 61.48 & 61.95 & 61.18 & $-6.70$ \\
EXAONE 3.5 7.8B & fin & KMMLU & 45.35 & 39.83 & 38.40 & 38.11 & 39.64 & $-5.71$ \\
EXAONE 3.5 7.8B & fin & KMMLU-fin & 51.24 & 44.11 & 44.74 & 44.97 & 46.09 & $-5.15$ \\
EXAONE 3.5 7.8B & fin & MMLU & 67.88 & 62.08 & 61.95 & 62.40 & 62.05 & $-5.83$ \\
\midrule
Qwen 3.5 9B & med & KMMLU & 58.98 & 56.67 & 57.16 & 56.31 & 58.05 & $-0.93$ \\
Qwen 3.5 9B & med & KMMLU-med & 61.35 & 58.02 & 57.97 & 58.73 & 57.90 & $-3.45$ \\
Qwen 3.5 9B & med & MMLU & 83.28 & 77.24 & 76.86 & 77.26 & 77.45 & $-5.83$ \\
Qwen 3.5 9B & fin & KMMLU & 58.98 & 56.74 & 57.40 & 57.78 & 59.01 & $+0.03$ \\
Qwen 3.5 9B & fin & KMMLU-fin & 61.43 & 58.38 & 58.81 & 59.20 & 60.09 & $-1.34$ \\
Qwen 3.5 9B & fin & MMLU & 83.28 & 77.21 & 76.84 & 77.00 & 77.36 & $-5.92$ \\
\midrule
Gemma 3 27B & med & KMMLU & 55.01 & 53.27 & 53.20 & 52.80 & 52.84 & $-2.17$ \\
Gemma 3 27B & med & KMMLU-med & 59.33 & 57.12 & 56.75 & 56.15 & 55.66 & $-3.67$ \\
Gemma 3 27B & med & MMLU & 77.88 & 74.83 & 74.09 & 74.41 & 74.48 & $-3.40$ \\
Gemma 3 27B & fin & KMMLU & 55.01 & 51.86 & 51.67 & 51.87 & 51.76 & $-3.25$ \\
Gemma 3 27B & fin & KMMLU-fin & 60.91 & 59.25 & 57.65 & 57.62 & 57.14 & $-3.77$ \\
Gemma 3 27B & fin & MMLU & 77.88 & 74.71 & 74.05 & 74.34 & 74.43 & $-3.45$ \\
\midrule
Llama 3.3 70B & med & KMMLU & 57.66 & 55.64 & 56.86 & 55.15 & 56.46 & $-1.20$ \\
Llama 3.3 70B & med & KMMLU-med & 61.15 & 56.63 & 58.34 & 56.79 & 57.58 & $-3.57$ \\
Llama 3.3 70B & med & MMLU & 83.37 & 82.19 & 82.42 & 82.28 & 82.27 & $-1.10$ \\
Llama 3.3 70B & fin & KMMLU & 57.66 & 56.06 & 55.97 & 55.83 & 56.25 & $-1.41$ \\
Llama 3.3 70B & fin & KMMLU-fin & 61.84 & 59.41 & 59.54 & 58.94 & 59.86 & $-1.98$ \\
Llama 3.3 70B & fin & MMLU & 83.37 & 82.10 & 82.03 & 81.92 & 82.09 & $-1.28$ \\
\bottomrule
\end{tabular}
\caption{Per-cell MCQA accuracy across all 9 models (54 rows: 9 models $\times$ 2 training domains $\times$ 3 domain-aligned metrics). LoRA cells: multi-seed mean over $n\!=\!2$ seeds; base models: single-seed. Domain-aligned subset only: medical-trained cells report KMMLU overall, KMMLU-medical, and MMLU; finance-trained cells report KMMLU overall, KMMLU-finance, and MMLU. Cross-aggregates in Table~\ref{tab:factorial} are computed from this table. $\Delta_{\text{2-base}}\!=\!\text{Exp~2}\!-\!\text{Base}$.}
\label{tab:app:all72}
\end{table*}